\documentclass[12pt]{article}

\usepackage[dvipsnames]{xcolor}
\usepackage{graphicx}
\usepackage{amsmath}        
\usepackage{amssymb}        
\usepackage{slashed}
\usepackage{bm}
\usepackage{here}
\usepackage{cite}
\def\mydate{July 15,  2026}
\def\ignore#1{{}}

\def\go{\rightarrow}
\def\dd{\partial}

\def\ep{{\epsilon}}

\def\KK{{\rm KK}}
\def\EM{{\rm EM}}

\def\max{{\rm max}}
\def\onehalf{\hbox{$\frac{1}{2}$}}

\def\la{\langle}
\def\ra{\rangle}

\def\Tr{{\rm Tr} \,}

\def\mymat#1#2{\begin{matrix}#1 \cr \noalign{\kern -2pt} #2\end{matrix}}

\def\mynoalign{\noalign{\kern 4pt}}
\def\mysnoalign{\noalign{\kern 3pt}}
\def\mytinynoalign{\noalign{\kern 2pt}}
\def\ignore#1{{}}

\makeatletter
\@addtoreset{equation}{section}
\makeatother

\begin{document}

\thispagestyle{empty}

%%%%% date and PREPRINT NUMBERS %%%%%%
{\small \noindent \mydate   \hfill }

\vskip 2.5cm

%%%%%%%%%%%%%%%%%%% TITLE %%%%%%%%%%%%%%%%%%
\baselineskip=35pt plus 1pt minus 1pt

\begin{center}
{\bf \Large Anomaly flow and anomaly cancellation}\\ %[12pt]
%{\bf \Large in gauge-Higgs unification}\\ %[12pt]
%{\bf \LARGE Neutrino oscillations and PMNS matrix}\\ %[12pt]
%{\bf \LARGE in gauge-Higgs unification}\\ %[12pt]
\end{center}

%%%%%%%%%%%%%%%% AUTHORS %%%%%%%%%%%%%%%%%%%%%%%

%\vspace{.0cm}
\baselineskip=22pt plus 1pt minus 1pt

\vskip 1.5cm

\begin{center}
\renewcommand{\thefootnote}{\fnsymbol{footnote}}
{\bf  Yutaka Hosotani\footnote[1]{hosotani@rcnp.osaka-u.ac.jp}}

\baselineskip=18pt plus 1pt minus 1pt

\vskip 10pt
{\small \it Research Center for Nuclear Physics, University of Osaka}\\
{\small \it Ibaraki, Osaka 567-0047, Japan}\\
%{\small \it  Ibaraki, Osaka 567-0047, Japan} \\

\end{center}

\vskip 2.cm
\baselineskip=18pt plus 1pt minus 1pt

\begin{abstract}
In gauge theory on 5D orbifolds  the magnitude of chiral anomalies of 4D gauge fields
changes with the value of the Aharonov-Bohm (AB) phase $\theta_H$  in the fifth dimension.
Anomaly  flows with the AB phase.  In particular in the Randall-Sundrum (RS) warped space
gauge couplings of 4D fermions depend on bulk mass parameters of 5D fermion multiplets.
We show that in the GUT-inspired $SO(5) \times U(1) \times SU(3)$  gauge-Higgs unification model in the 
RS  warped  space the total anomalies
including contributions of all Kaluza-Klein excited modes of fermions become universal, being independent of the values
of the bulk mass parameters of fermions and expressed in terms of the values of $W$ and $Z$ wave functions  
at the ultraviolet and infrared branes in the RS space.  It is shown that  cancellation of gauge anomalies is achieved in each generation
even for   $\theta_H \not= 0$.
\end{abstract}

\newpage

\baselineskip=20pt plus 1pt minus 1pt
\parskip=0pt

\section{Introduction}

In 4D gauge theory cancellation of gauge anomalies is necessary for the consistency of theory.
In the standard model (SM) of  $SU(3)_C \times SU(2)_L \times U(1)_Y$ gauge theory  the cancellation is achieved 
in each generation\cite{Adler1969,  BellJackiw1969, Bouchiat1972, GrossJackiw1972}.
Although the SM has been successful in describing  phenomena at low energies, it has a gauge hierarchy problem.  
% when embedded in a larger theory such as grand unification.
Gauge-Higgs unification (GHU) models have been proposed to solve the gauge hierarchy problem by
unifying the Higgs boson  and gauge fields in five dimensions.
The  gauge symmetry $SU(2)_L \times U(1)_Y$ is dynamically broken by an Aharonov-Bohm (AB)  phase, 
$\theta_H$, in the fifth dimension.
The Higgs boson appears as a 4D fluctuation mode of  $\theta_H$, and 
its  mass is  generated by quantum effects of the AB phase 
$\theta_H$\cite{Hosotani1983, Davies1988, Hosotani1989, Davies1989,  
Hatanaka1998, Hatanaka1999,  Antoniadis2001, Takenaga2002, Kubo2002,   BurdmanNomura2003,  Csaki2003,  
Scrucca2003, Cacciapaglia2006,  Medina2007, HOOS2008, Serone2010,  Yoon2018b,
GUTinspired2019a}.  
In this paper we reexamine the issue of anomaly cancellation in GHU models.
%HetrickHo1989, McLachlan1990, FHHOS2013, %. ACP,  FCNC2020a, GUTinspired2020b

Among various GHU models the $SO(5)\times U(1) \times SU(3)$ GHU in the Randall-Sundrum (RS) 
warped  space\cite{GUTinspired2019a},   which is inspired by $SO(11)$ gauge-Higgs grand unification 
models \cite{SO11GHGU},  is a promising realistic model.
The grand unified theory (GUT)-inspired GHU 
gives nearly the same phenomenology at low energies as in the SM.  
It has been shown recently that the Cabibbo-Kobayashi-Maskawa (CKM) matrix in the quark sector is 
reproduced \cite{CKM2025}  and that neutrino oscillations are explained and  the Pontecorvo-Maki-Nakagawa-Sakata
(PMNS) matrix  in the normal ordering with $\delta_{CP} = \pi$ naturally arises\cite{neutrino2026}.
Further it predicts distinct deviation from the SM in forward-backward asymmetry in fermion pair production 
in electron-positron ($e^- e^+$)  collisions  and smaller Higgs cubic and quartic self-couplings than those 
in the SM\cite{YHbook}.

One of the characteristics of GHU models in the RS space is that  gauge couplings of quarks and leptons
to $W$ and $Z$ bosons at $\theta_H \not= 0$ are  not exactly the same as those in the SM,
showing tiny deviation from those in the SM.
Furthermore those couplings depend not only on $\theta_H$, but also on bulk mass parameters of
fermion multiplets.  It implies that if only contributions of quarks and leptons to gauge anomalies, 
namely those of the lowest modes of their Kaluza-Klein (KK) towers, were taken into account,
then gauge anomalies would not be cancelled in a rigorous sense. 
From the viewpoint of effective 4D gauge theory this would cause a serious danger for theoretical consistency. 

GHU models on an orbifold such as the RS space exhibit the phenomenon of anomaly flow.
The magnitude of gauge anomalies varies as the AB phase $\theta_H$ changes.  
Previously this phenomenon was investigated in an $SU(2)$ toy model \cite{AnomalyFlow1, AnomalyFlow2, AnomalyFlow3}.
It was shown there that one has to incorporate, for gauge anomalies, contributions of all KK excited modes of fermions
running along internal triangular loops.  With all those contributions taken into account, total anomalies become universal.
We note that the phenomenon of anomaly flow is different from that of anomaly inflow \cite{CallanHarvey1985}.
Further it was shown in Ref.\ \cite{AnomalyFlow2} that the magnitude of anomalies is expressed in terms of the values of
the wave functions of gauge fields on the ultraviolet (UV) and infrared (IR) branes.
In this paper we show that  the magnitude of gauge anomalies in the GUT-inspired GHU model is expressed
in terms of the values of the  $W$ and $Z$ wave functions  on the UV and IR branes, holographic formulas
for anomalies being established.  With these formulas cancellation of gauge anomalies
in the GUT-inspired GHU is achieved.

In Section 2 the GUT-inspired $SO(5) \times U(1)_X \times SU(3)_C$ GHU model  is described.  
With a general $\theta_H$ mass spectra and wave functions of gauge field  and fermion 
multiplets are given.  In Section 3 gauge couplings of all KK modes of fermions are given.
In Section 4 chiral anomalies associated with gauge bosons in the SM are expressed in terms of 
gauge couplings derived in Section 3.  In Section 5 holographic formulas for chiral anomalies are 
derived.  The validity of the formulas is checked by numerical evaluations of gauge couplings of 
all KK fermions as well.  The $\theta_H$-dependence of anomalies is also investigated.
With the use of holographic formulas for anomalies we show the cancellation of gauge anomalies
in the GUT-inspired GHU  in Section 6.  A brief summary is given in Section 7.
In Appendix A basis functions used to express wave functions of gauge and fermion fields in the RS space are given.
In Appendix B  wave functions of fermions with a vanishing bulk mass parameter $c=0$ are explained.
In Appendix C numerical results for establishing universality of anomaly flow in the GUT-inspired  GHU model 
are summarized.
 
 \section{GUT-inspired GHU} 
 
We analyze gauge anomaly in the GUT inspired $SO(5) \times U(1)_X \times SU(3)_C (\equiv {\cal G})$ 
GHU  \cite{GUTinspired2019a, YHbook}.  
The model is defined in the RS warped space with the metric \cite{RS1}
\begin{align}
ds^2= G_{MN} \,  dx^M dx^N =
e^{-2\sigma(y)} \eta_{\mu\nu}dx^\mu dx^\nu+dy^2,
\label{RSmetric1}
\end{align}
where $M,N=0,1,2,3,5$, $\mu,\nu=0,1,2,3$, $y=x^5$, $\eta_{\mu\nu}=\mbox{diag}(-1,+1,+1,+1)$,
$\sigma(y)=\sigma(y+ 2L)=\sigma(-y)$, and $\sigma(y)=ky$ for $0 \le y \le L$.
In terms of the conformal coordinate $z=e^{ky}$ ($0 \le y \le L$, $1\leq z\leq z_L=e^{kL}$)
the metric becomes
\begin{align}
ds^2=  \frac{1}{z^2} \bigg(\eta_{\mu\nu}dx^{\mu} dx^{\nu} + \frac{dz^2}{k^2}\bigg)~ .
\label{RSmetric-2}
\end{align}
The bulk region $0<y<L$ is anti-de Sitter spacetime 
with a cosmological constant $\Lambda=-6k^2$, which is sandwiched by the
UV brane at $y=0$  and the IR brane at $y=L$.  
The warp factor $z_L$ is large.
The KK mass scale is given by $m_{\rm KK}=\pi k/(z_L-1) \simeq \pi kz_L^{-1}$.
Typical values are $z_L \sim 10^{11}$  and $m_\KK  \sim 13\,$TeV.

Gauge fields 
$A_M^{SO(5)}$,  $A_M^{U(1)_X}$ and $A_M^{SU(3)_C}$ of $SO(5) \times U(1)_X \times SU(3)_C$
with gauge couplings  $g_A$,  $g_B$ and $g_S$
satisfy the orbifold boundary conditions (BCs)
\begin{align}
\begin{pmatrix} A_\mu \cr  A_{y} \end{pmatrix} (x,y_j-y) &=
P_{j} \begin{pmatrix} A_\mu \cr  - A_{y} \end{pmatrix} (x,y_j+y)P_{j}^{-1}
\quad (j=0,1), \cr
\noalign{\kern 5pt}
(y_0, y_1) &= (0, L) .
\label{BC-gauge1}
\end{align}
Here   $P_0=P_1 = P_{\bf 5}^{SO(5)} =\mbox{diag} (I_{4},-I_{1} )$ for $A_M^{SO(5)}$ in the vector 
representation and  $P_0=P_1 = P_{\bf 4}^{SO(5)} =\mbox{diag} (I_{2},-I_{2} )$  for 
$A_M^{SO(5)}$ in the spinorial  representation.    $P_0=P_1= 1$ for $A_M^{U(1)_X}$ and $A_M^{SU(3)_C}$.
The orbifold BCs break $SO(5)$ to $SO(4) \simeq SU(2)_L \times SU(2)_R$.
In the following we write $A_M^{SO(5)} = A_M$ and  $A_M^{U(1)_X} = B_M$ unless confusion arises.

 The 4D Higgs boson doublet $\phi_H(x)$ is  contained in  $A_z = (kz)^{-1} A_y$;
\begin{align}
A_z^{(j5)} (x, z) &= \frac{1}{\sqrt{k}} \, \phi_j (x) u_H (z) + \cdots , ~~
u_H (z) = \sqrt{ \frac{2}{z_L^2 -1} } \, z ~, \cr
\noalign{\kern 5pt}
\phi_H(x) &= \frac{1}{\sqrt{2}} \begin{pmatrix} \phi_2 + i \phi_1 \cr \phi_4 - i\phi_3 \end{pmatrix} .
\label{4dHiggs}
\end{align}
Without loss of generality one can assume $\la \phi_1 \ra , \la \phi_2 \ra , \la \phi_3 \ra  =0$ and  
$\la \phi_4 \ra \not= 0$.  
The AB phase $\theta_H$ in the fifth dimension is given by $\la \phi_4 \ra  = \theta_H f_H$ where
\begin{align}
&P \exp \bigg\{ i g \oint dy \, A_y \bigg\}
\sim \exp \bigg\{ i \Big(  \theta_H + \frac{H(x)}{f_H} \Big) \, 2 \, T^{(45)}\bigg\} ~, \cr
\noalign{\kern 5pt}
&f_H  = \frac{2}{g_w} \sqrt{ \frac{k}{L(z_L^2 -1)}} ~~,~ g_w = \frac{g_A}{\sqrt{L}} ~.
\label{fH1}
\end{align}
Physics is periodic in $\theta_H$ with a period $2 \pi$.  
(Note that eigenvalues of $2 T^{(45)}_{\rm sp}$ in the spinorial representation are $\pm 1$.)
The neutral Higgs field $H(x)$ is a four-dimensional fluctuation mode of $\theta_H$.

The matter field content of the model is summarized in Table \ref{Table:GHUmatter}.
The orbifold BCs for fermion fields in the bulk are
\begin{align}
\Psi_{({\bf 3,4})}^{q , \alpha} (x, y_j - y)  &= 
- P_{\bf 4}^{SO(5)} \gamma^5 \Psi_{({\bf 3,4})}^{q, \alpha} (x, y_j + y) ~, \cr
\Psi_{({\bf 1,4})}^{\ell , \alpha} (x, y_j - y) &= 
- P_{\bf 4}^{SO(5)} \gamma^5 \Psi_{({\bf 1,4})}^{\ell , \alpha} (x, y_j + y) ~, \cr
\Psi_{({\bf 3,1})^\pm }^{q, \alpha}  (x, y_j - y)  &=
\mp \gamma^5 \Psi_{({\bf 3,1})^\pm}^{q, \alpha}  (x, y_j + y) ~, \cr
&\alpha = 1 , 2 , 3 , 
\label{quarkleptonBC1}
\end{align}
for quark and lepton multiplets  and 
\begin{align}
&\Psi^{F_q, \alpha}_{({\bf 3},{\bf 4})}  (x, y_j - y) = 
(-1)^j  P_{\bf 4}^{SO(5)}  \gamma^5 \Psi^{F_q, \alpha}_{({\bf 3},{\bf 4})} (x, y_j + y) ~, \cr
&\Psi^{F_\ell, \alpha}_{({\bf 1},{\bf 4})}  (x, y_j - y) = 
(-1)^j  P_{\bf 4}^{SO(5)}  \gamma^5 \Psi^{F_\ell, \alpha}_{({\bf 1},{\bf 4})} (x, y_j + y) ~, \cr
&\Psi^{V, \beta}_{({\bf 1},{\bf 5})^\pm}  (x, y_j - y) =
\mp P_{\bf 5}^{SO(5)} \gamma^5 \Psi^{V, \beta}_{({\bf 1},{\bf 5})^\pm} (x, y_j + y) ~, \cr
&\qquad
 \alpha = 1, \cdots, n_F , ~~~   \beta = 1, \cdots, n_V ,
\label{DFBC1}
\end{align}
for dark fermion multiplets.
The kinetic part of the action in the bulk of each fermion multiplet $\Psi$ takes the form of 
\begin{align}
&\int d^5 x \, \sqrt{-G}  ~ \overline{\Psi} \, {\cal D} (c) \Psi ~, \cr
\noalign{\kern 5pt}
&{\cal D} (c)   = \Gamma^a {e_a}^M \Big( D_M 
+ \frac{1}{8} \omega_{bcM} [\Gamma^b, \Gamma^c]  \Big) - c \,  \sigma' (y) ~,  \cr
\noalign{\kern 5pt} 
&D_M = \dd_M  - i g_A   A_M  -i  g_B  Q_X B_M - i g_C A_M^{SU(3)_C} ~.
\label{fermionaction1}
\end{align}
Here $\omega_{bc} = \omega_{bcM} d x^M$ is the spin-connection 1-form.  $Q_X$ and $c$ are
$U(1)_X$ charge and  bulk mass parameter of the $\Psi$ field.

%%%%%%%%%%%%%%%

\begin{table}[bh]
\renewcommand{\arraystretch}{1.4}
\begin{center}
\caption{The matter field content in the GUT-inspired  GHU model.
The $(SU(3)_C, SO(5))_{U(1)_X}$ content of each field is shown in the last column.}
{\vskip 10pt}
%\vspace 10pt
{\begin{tabular}{ccc}
\hline 
in the bulk &quark $\Psi^q$
&$({\bf 3}, {\bf 4})_{\frac{1}{6}} ~ ({\bf 3}, {\bf 1})_{-\frac{1}{3}}^+ 
    ~ ({\bf 3}, {\bf 1})_{-\frac{1}{3}}^-$\\
%\cline{2-3}
$(1 \le z  \le z_L)$ &lepton $\Psi^\ell$
&$\strut ({\bf 1}, {\bf 4})_{-\frac{1}{2}}$ \\
%\cline{2-3}
&dark fermion $\Psi^{F_q}, \Psi^{F_\ell}$ & $({\bf 3}, {\bf 4})_{\frac{1}{6}} \, ,  \, ({\bf 1}, {\bf 4})_{-\frac{1}{2}}$  \\
% &dark fermion $\Psi^{F_\ell}$ & $({\bf 1}, {\bf 4})_{-\frac{1}{2}}$  \\
&dark fermion $\Psi^V$ & $({\bf 1}, {\bf 5})_{0}^+ ~ ({\bf 1}, {\bf 5})_{0}^-$  \\
\hline 
on the UV brane
&Majorana fermion $\hat \chi$ &$({\bf 1}, {\bf 1})_{0} $ \\
%&{\small (Majorana)}
%&$({\bf 1}, [{\bf 1,2}])_{\frac{1}{2}, -\frac{1}{2}, -\frac{3}{2}}$ \\
%\cline{2-3}
(at $z=1$) &brane scalar $\hat \Phi$ &$({\bf 1}, {\bf 4})_{\frac{1}{2}} $ \\
\hline 
\end{tabular}
}
\label{Table:GHUmatter}
\end{center}
\end{table}

%%%%%%%%%%%%%%%%

\begin{figure}[tbh]
\centering
\includegraphics[height=80mm]{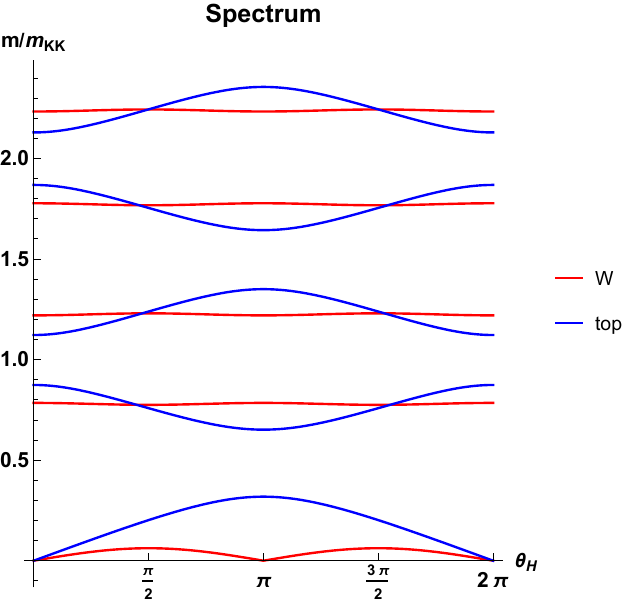}
\caption{The mass spectra of the $W$ and top quark KK towers in the RS space in units of $m_\KK = 13\,$TeV.
The observed $m_W$  and $m_t$ are reproduced at $\theta_H=0.1$.}
\label{fig:mWtop}
\end{figure}

Parity assignment and $G_{22}=SU(2)_L\times SU(2)_R$ content of quark-lepton multiplets 
with their names are tabulated in Table \ref{Table:quarklepton}.
A consistent models is obtained for small $\theta_H$.  
As typical values let us take  $\theta_H = 0.1$ and KK mass scale $m_\KK = \pi k /(z_L -1) = 13\,$TeV.
The parameters of the model are fixed such that $m_Z$, quark-lepton masses, fine structure constant $\alpha_\EM$, 
QCD coupling $\alpha_S$, Fermi constant $G_F$, weak mixing angle $\sin^2 \theta_W$, 
and Higgs boson mass $m_H$  are reproduced.
With the mixing in the mass matrix of $\Psi_{({\bf 3,1})^\pm}^{q, \alpha}$ fields in the quark sector and
the mixing in Majorana mass terms for $\hat \chi^\alpha$ fields, both of the  CKM matrix and
PMNS matrix are obtained \cite{CKM2025, neutrino2026}.
The bulk mass parameters of $\Psi_{({\bf 3,4})}^{q, \alpha}$ and $\Psi_{({\bf 1,4})}^{\ell, \alpha}$
are $(c_{u}, c_{c}, c_{t}) = (-0.85912,  -0.71913,  - 0.27455)$ and
$ (c_e, c_\mu, c_\tau) = (-1.00684, -0.79302, -0.67539)$
for $\theta_H=0.1$ and $m_\KK=13\,$TeV.
The mass spectra of $W$ and top quark KK towers are depicted in Fig.\ \ref{fig:mWtop}.

\begin{table}[tbh]
\renewcommand{\arraystretch}{1.2}
\begin{center}
\caption{Parity assignment $(P_0, P_1)$ of quark-lepton multiplets in the bulk is shown.
%The corresponding names adopted in Ref.~\cite{Furui2016} are 
%listed in the last column for the first generation.
%$G_{3221}=SU(3)_C\times SU(2)_L\times SU(2)_R\times U(1)_X$.
In the second column $\big( SU(3)_C, SO(5) \big)_{U(1)_X}$ content is shown.
In the third column $G_{22} =SU(2)_L\times SU(2)_R$ content is shown.
}
\vskip 10pt
\begin{tabular}{cccccc}
%\begin{tabular}{|c|c|c|c|c|c|}
\hline \hline 
Field & ${\cal G}$ & $G_{22}$ &Left-handed &Right-handed &Name\\
\hline
$\Psi_{({\bf 3,4})}^{q, \alpha}$ &$({\bf 3,4})_{\frac{1}{6}}$ &$\, [{\bf 2} , {\bf  1}] \,$
&$(+,+)$ &$(-,-)$ &$u ~~~ c~~~ t$\\
%&$(+,+)$ &$(-,-)$ &$\begin{matrix} u~ & c & ~t \cr d~ & s & ~b\end{matrix}$\\
&&&&& $d ~~~ s ~~~ b$\\
%\cline{3-6}
&&$[{\bf 1} , {\bf  2}]$ 
&$(-,-)$ &$(+,+)$ &$u' ~~~ c'~~~ t'$\\
%&$(-,-)$ &$(+,+)$ &$\begin{matrix} u'  & c' & t' \cr d' & s' & b' \end{matrix}$\\
&&&&& $d' ~~~ s'~~~ b'$\\
%\hline
$\Psi_{({\bf 3,1})^\pm}^{q, \alpha}$ &$({\bf 3,1})_{-\frac{1}{3}}$ 
&$[{\bf 1} , {\bf  1}]$
&$(\pm ,\pm )$ &$(\mp , \mp )$ &$D^{\pm}_d ~ D^{\pm}_s ~ D^{\pm}_b$\\
%\hline %\hline
\noalign{\kern 5pt}
\hline
\noalign{\kern 5pt}
$\Psi_{({\bf 1,4})}^{\ell, \alpha}$ &$({\bf 1,4})_{- \frac{1}{2}}$ &$\, [{\bf 2} , {\bf  1}] \,$
&$(+,+)$ &$(-,-)$ &$\nu_e \,~  \nu_\mu ~\,  \nu_\tau$\\
&&&&& $e ~~~ \mu ~~~ \tau$\\
%\cline{3-6}
&&$[{\bf 1} , {\bf  2}]$ 
&$(-,-)$ &$(+,+)$ &$\nu_e' \,~  ~ \nu_\mu' ~\,  \nu_\tau'$\\
&&&&& $e' ~~~ \mu' ~~ \tau'$\\
\hline

$\hat \chi^\alpha$ &$({\bf 1,1})_0$ 
&$[{\bf 1} , {\bf  1}]$
&$\cdots$ &$\cdots$ &$\eta_e ~\, \eta_\mu ~\, \eta_\tau$\\
\hline \hline
\end{tabular}
\label{Table:quarklepton}
\end{center}
\end{table}

%%%%%%%%%%%%%%%%%%%%
%%%%%%%%%%%%%%%%%%%

Gauge couplings of quarks and leptons to $W$ and $Z$ bosons are determined 
by integrating over the overlaps of respective wave functions in the fifth dimension.
Manipulations are simplified in the twisted gauge %\cite{Falkowski2007, HS2007}
defined by  %an $SO(5)$ large gauge transformation
\begin{align}
&\tilde A_M (x,z) = \Omega A_M \Omega^{-1} 
- \frac{i}{g_A} \, \Omega \,\dd_M \Omega^{-1} ~, \cr
\noalign{\kern 5pt}
& \Omega (z)  = \exp \Big\{ i \theta (z) T^{45} \Big\}  ~,~~
\theta (z) = \theta_H \, \frac{z_L^2 - z^2}{z_L^2 - 1} ~, 
\label{twisted1}
\end{align}
where $T^{jk}$'s are $SO(5)$ generators and  
$A_M = 2^{-1/2} \sum_{1 \le j<k \le 5} A_M^{(jk)} T^{jk}$.
In the twisted gauge the background field vanishes ($\tilde \theta_H = 0$),
whereas boundary conditions at the UV brane are modified.
Boundary conditions at the IR brane remain  the same as in the original gauge.  
Quantities in the twisted gauge are denoted by the tilde sign $\tilde{~}$.

It is convenient to adopt the $SO(5)$ basis of $\{ SU(2)_L, SU(2)_R, SO(5)/SO(4) \}$;
\begin{align}
&\{ T^\alpha ; \alpha = 1 \sim 10 \} = 
\{ T^{a_L} , T^{a_R} \, (a = 1 , 2, 3) , ~ T^{\hat p} \, (p=1 \sim 4)  \} , \cr
\noalign{\kern 5pt}
&\begin{pmatrix} T^{a_L}  \cr T^{a_R} \end{pmatrix} = \frac{1}{4} \ep^{abc} T^{bc}  \pm \frac{1}{2} T^{a4} , ~~
T^{\hat p}  = \frac{1}{\sqrt{2}} \, T^{p5} ~.
\label{so5basis}
\end{align}
%where $a,b,c = 1 \sim 3$ and $p= 1 \sim 4 $.
$SO(5)$ gauge fields are expressed as
\begin{align}
&A_M = \sum_{\alpha=1}^{10} A_M^\alpha T^\alpha ~,  \cr
\noalign{\kern 5pt}
&\begin{pmatrix} A_M^{a_L} \cr A_M^{a_R} \end{pmatrix} = 
\frac{1}{\sqrt{2}}  \Big( \frac{1}{2} \ep^{abc} A_M^{bc} \pm A_M^{a4} \Big) , ~~
A_M^{\hat p} = A_M^{p5} ~. \
\label{so5A}
\end{align}

The $W$ boson field, $W_\mu (x)$,  is contained in 
$\sum_{a=1}^2 ( A_\mu^{a_L}  T^{a_L} + A_\mu^{a_R}  T^{a_R} + A_\mu^{\hat a} T^{\hat a} )$.
In the twisted gauge its KK expansion  in the region $1 \le z \le z_L$ is given by
\begin{align}
&\frac{1}{\sqrt{2 k}} \begin{pmatrix} \tilde A_\mu^{1_L} + i \tilde A_\mu^{2_L} \cr
\tilde A_\mu^{1_R} + i \tilde A_\mu^{2_R} \cr 
\tilde A_\mu^{\hat 1} + i \tilde A_\mu^{\hat 2} \end{pmatrix} 
=  \sum_{n=0}^\infty W_\mu^{ (n)} (x) \begin{pmatrix}  \tilde h^L_{W^{(n)}} (z) \cr \mysnoalign
 \tilde h^R_{W^{(n)}} (z) \cr  \mysnoalign  \tilde{\hat h}_{W^{(n)}} (z) \end{pmatrix} + \cdots , \cr
\noalign{\kern 5pt}
& \begin{pmatrix}  \tilde h^L_{W^{(n)}} (z) \cr \mysnoalign
 \tilde h^R_{W^{(n)}} (z) \cr  \mysnoalign  \tilde{\hat h}_{W^{(n)}} (z) \end{pmatrix} =
 \frac{1}{\sqrt{2 \, r_{W^{(n)}}}}
\begin{pmatrix} (1 + c_H) \, C(z; \lambda_{W^{(n)}}) \cr 
(1 - c_H) \, C(z; \lambda_{W^{(n)}})  \cr  
\sqrt{2} \, s_H \check S(z; \lambda_{W^{(n)}}) \end{pmatrix}  , \cr
\noalign{\kern 5pt}
&\quad c_H = \cos \theta_H ~, ~~ s_H = \sin \theta_H ~, 
\label{WbosonWave1}
\end{align}
where the spectrum $m_{W^{(n)}} = k \lambda_{W^{(n)}}$ is determined by
\begin{align}
&2 S C' (1; \lambda_{W^{(n)}}) +   \lambda_{W^{(n)}} \sin^2 \theta_H  =0 ~.
\label{Wspectrum1}
\end{align}
% $2 S C' (1; \lambda_{W^{(n)}}) +s_H^2  \lambda_{W^{(n)}}  =0$.
$C(z; \lambda)$,  $S(z; \lambda)$, and $\check S(z; \lambda)$ are defined in Eq.\  (\ref{functionA1}).
The normalization factor $r_{W^{(n)}}$ is  determined by
\begin{align}
&\int_1^{z_L}  \frac{dz}{z} \Big\{ ( |\tilde h^L_{W^{(n)}} |^2 + |\tilde h^R_{W^{(n)}} |^2 
+ | \tilde{\hat h}_{W^{(n)}} |^2 \Big\} = 1 ~.
\label{Wnormalization}
\end{align}
The $W$ boson field is $W_\mu (x) = W_\mu^{(0)}  (x)$.
Wave functions in the original gauge are given by
\begin{align}
\begin{pmatrix}  h^L_{W^{(n)}} (z) \cr \mysnoalign h^R_{W^{(n)}} (z) \end{pmatrix} 
&=  \frac{1 \pm \cos \theta (z)}{2}  \,  \tilde h^L_{W^{(n)}} (z) 
+  \frac{1 \mp \cos \theta (z)}{2}  \,  \tilde h^R_{W^{(n)}} (z) 
\pm \frac{\sin \theta(z)}{\sqrt{2}} \, \tilde{\hat h}_{W^{(n)}} (z) ~, \cr
\noalign{\kern 5pt}
{\hat h}_{W^{(n)}} (z)  &= -  \frac{\sin \theta(z)}{\sqrt{2}} \,  \tilde h^L_{W^{(n)}} (z) 
+ \frac{\sin \theta(z)}{\sqrt{2}} \,  \tilde h^R_{W^{(n)}} (z)  + \cos \theta (z) \, \tilde{\hat h}_{W^{(n)}} (z) ~.
\label{WbosonWave2}
\end{align}
Here $\theta (z)$ is given in Eq.\  (\ref{twisted1}).

Similarly the spectrum of the $Z$ boson tower, $\{ m_{Z^{(n)}} = k \lambda_{Z^{(n)}}  \}$,  is determined by 
\begin{align}
&2 S C' (1; \lambda_{Z^{(n)}}) +  \lambda_{Z^{(n)}} \frac{\sin^2 \theta_H}{\cos^2 \theta_W^0}=0 ~, \cr
\noalign{\kern 5pt}
&\sin \theta_W^0  = \frac{g_B}{\sqrt{g_A^2 + 2 g_B^2}} ~.
\label{Zspectrum1}
\end{align}
The $Z$ boson field is contained in 
$A_\mu^{3_L}$, $ A_\mu^{3_R} $, $A_\mu^{\hat 3} $, and $B_\mu $.
Wave functions of the $Z$-boson tower in the twisted gauge are given by \cite{YHbook}
\begin{align}
&\frac{1}{\sqrt{k}} \begin{pmatrix} \tilde A_\mu^{3_L}  \cr \mytinynoalign \tilde A_\mu^{3_R}\cr 
\mytinynoalign  \tilde A_\mu^{\hat 3}  \cr B_\mu  \end{pmatrix} 
= \sum_{n=0}^\infty Z_\mu^{ (n)} (x) \begin{pmatrix}  \tilde h^L_{Z^{(n)}} (z) \cr \mysnoalign
\tilde  h^R_{Z^{(n)}} (z) \cr  \mysnoalign    \tilde {\hat h}_{Z^{(n)}} (z) \cr  \mysnoalign h^B_{Z^{(n)}} (z) \end{pmatrix} + \cdots , \cr
\noalign{\kern 5pt}
& \begin{pmatrix}  \tilde h^L_{Z^{(n)}} (z) \cr \mysnoalign
\tilde  h^R_{Z^{(n)}} (z) \cr  \mysnoalign    \tilde {\hat h}_{Z^{(n)}} (z) \cr  \mysnoalign h^B_{Z^{(n)}} (z) \end{pmatrix}  = 
  \begin{pmatrix}  \tilde h^{L, su2}_{Z^{(n)}} (z) \cr \mysnoalign
\tilde  h^{R, su2}_{Z^{(n)}} (z) \cr  \mysnoalign    \tilde {\hat h}^{su2}_{Z^{(n)}} (z) \cr   0 \end{pmatrix}
 - \sin \theta_W^0  \begin{pmatrix} \sin \theta_W^0 \cr   \sin \theta_W^0 \cr     0 \cr 
\sqrt{1 - 2 \sin^2 \theta_W^0} \end{pmatrix} h^{em}_{Z^{(n)}} (z) , \cr
\noalign{\kern 5pt}
&\hskip 0.7cm
 \begin{pmatrix}  \tilde h^{L, su2}_{Z^{(n)}} (z) \cr \mysnoalign
 \tilde h^{R, su2}_{Z^{(n)}} (z) \cr  \mysnoalign    \tilde {\hat h}^{su2}_{Z^{(n)}} (z)  \end{pmatrix} =
\frac{1}{\sqrt{ \, 2 \, r_{Z^{(n)}} }} 
\begin{pmatrix} (1 + c_H) \,C(z, \lambda_{Z^{(n)}} ) \cr 
(1 -c_H) \,C(z, \lambda_{Z^{(n)}} ) \cr  
  \sqrt{2}s_H  \check S (z, \lambda_{Z^{(n)}}  ) \end{pmatrix} , \cr
\noalign{\kern 5pt}
&\hskip 1.cm
h^{em}_{Z^{(n)}} (z)  = \sqrt{\frac{2}{r_{Z^{(n)}}}} \, C(z, \lambda_{Z^{(n)}} )~.
\label{ZbosonWave1}
\end{align}
Wave functions are normalized by
\begin{align}
&\int_1^{z_L}  \frac{dz}{z} \Big\{ ( |\tilde h^L_{Z^{(n)}} |^2 + |\tilde h^R_{Z^{(n)}} |^2 + | \tilde {\hat h}_{Z^{(n)}} |^2  
+  |h^B_{Z^{(n)}} |^2 \Big\} = 1 ~.
\label{Znormalization}
\end{align}
The $Z$ boson field is $Z_\mu (x) = Z_\mu^{(0)}  (x)$.
Wave functions in the original gauge are given by
\begin{align}
& \begin{pmatrix} h^L_{Z^{(n)}} (z) \cr \mysnoalign
 h^R_{Z^{(n)}} (z) \cr  \mysnoalign   {\hat h}_{Z^{(n)}} (z) \cr  \mysnoalign h^B_{Z^{(n)}} (z) \end{pmatrix}  = 
  \begin{pmatrix}  h^{L, su2}_{Z^{(n)}} (z) \cr \mysnoalign
 h^{R, su2}_{Z^{(n)}} (z) \cr  \mysnoalign   {\hat h}^{su2}_{Z^{(n)}} (z) \cr   0 \end{pmatrix}
 - \sin \theta_W^0  \begin{pmatrix} \sin \theta_W^0 \cr   \sin \theta_W^0 \cr     0 \cr 
\sqrt{1 - 2 \sin^2 \theta_W^0} \end{pmatrix} h^{em}_{Z^{(n)}} (z) .
\label{ZbosonWave2}
\end{align}
Here $(h^{L, su2}_{Z^{(n)}} , h^{R, su2}_{Z^{(n)}} , {\hat h}^{su2}_{Z^{(n)}} )$ is related to
$(\tilde h^{L, su2}_{Z^{(n)}} , \tilde h^{R, su2}_{Z^{(n)}} , \tilde{\hat h}^{su2}_{Z^{(n)}} )$,
as in the formula (\ref{WbosonWave2}), 
by %the formula (\ref{WbosonWave2}) where $W^{(n)}$ is replaced by $Z^{(n)}$.
\begin{align}
\begin{pmatrix}  h^{L, su2}_{Z^{(n)}} (z) \cr \mysnoalign h^{R, su2}_{Z^{(n)}} (z) \end{pmatrix} 
&=  \frac{1 \pm \cos \theta (z)}{2}  \,  \tilde h^{L, su2}_{Z^{(n)}}  (z) 
+  \frac{1 \mp \cos \theta (z)}{2}  \,  \tilde h^{R, su2}_{Z^{(n)}} (z) 
\pm \frac{\sin \theta(z)}{\sqrt{2}} \, \tilde{\hat h}^{su2}_{Z^{(n)}} (z) ~, \cr
\noalign{\kern 5pt}
{\hat h}^{su2}_{Z^{(n)}} (z)  &= -  \frac{\sin \theta(z)}{\sqrt{2}} \,  \tilde h^{L, su2}_{Z^{(n)}}  (z) 
+ \frac{\sin \theta(z)}{\sqrt{2}} \,  \tilde h^{R, su2}_{Z^{(n)}} (z)  + \cos \theta (z) \, \tilde{\hat h}^{su2}_{Z^{(n)}} (z) ~.
\label{ZbosonWave3}
\end{align}
The $U(1)_\EM$ part $h^{em}_{Z^{(n)}} (z)$ is unchanged. 
Note that $h^{L, su2}_{Z^{(n)}} (z)  + h^{R, su2}_{Z^{(n)}} (z)  = h^{em}_{Z^{(n)}} (z) $.

In this paper we restrict ourselves to the case in which there is no flavor mixing in quarks and leptons, 
to make clear  the argument of anomaly flow and anomaly cancellation.  
The extension to a general case with flavor mixing is straightforward.
Up, charm, and top quarks are zero modes contained  in $\Psi_{({\bf 3,4})}^{\alpha}$.
The mass spectrum $m_{q^{(n)}} = k \lambda_{q^{(n)}}$  ($q= u, c, t$) is determined by
\begin{align}
&S_L (1;  \lambda_{q^{(n)}}, c_q)  S_R (1;  \lambda_{q^{(n)}}, c_q) +\sin^2\onehalf \theta_H =0 ~,
%&S_L (1, \lambda, c_q)  S_R (1, \lambda, c_q) +\sin^2\frac{\theta_H}{2}=0 ~.
\label{Up-quark-mass1}
\end{align}
where $S_{L/R} (z, \lambda, c)$ and $C_{L/R} (z, \lambda, c)$ are given by (\ref{functionA2}).
The lowest modes $u^{(0)}$, $c^{(0)}$ and $t^{(0)}$ are $u$, $c$, and $t$ quarks.
With $m_u$, $m_c$ and $m_t$ given, the corresponding bulk mass parameters $c_u$, $c_c$ and $c_t$ are fixed.
Let us define $\check \Psi = z^{-2} \Psi$.
The KK expansion of 5D $u (x,z)$ and $u' (x,z)$ fields is given,  in the twisted gauge, by
\begin{align}
& \begin{pmatrix} \tilde{\check u} \cr \tilde {\check u}' \end{pmatrix} = 
\sqrt{k} \sum_{n=0}^\infty \bigg\{ u^{(n)}_L (x) \begin{pmatrix} \tilde f^{u^{(n)}}_L (z) \cr \tilde g^{u^{(n)}}_L (z) \end{pmatrix}
+ u^{(n)}_R (x) \begin{pmatrix} \tilde f^{u^{(n)}}_R (z) \cr \tilde g^{u^{(n)}}_R (z) \end{pmatrix} \bigg\} , \cr
\noalign{\kern 5pt}
&\quad
\begin{pmatrix} \tilde f^{u^{(n)}}_L (z) \cr \tilde g^{u^{(n)}}_L (z) \end{pmatrix}=  \frac{1}{\sqrt{r_{u^{(n)} L}}}
\begin{pmatrix}\bar c_H C_L (z, \lambda_{u^{(n)}}, c_u) \cr
\noalign{\kern 5pt}
- i\bar s_H  \check S_L (z, \lambda_{u^{(n)}}, c_u) \end{pmatrix}, \cr
\noalign{\kern 5pt}
&\quad
\begin{pmatrix} \tilde f^{u^{(n)}}_R (z) \cr \tilde g^{u^{(n)}}_R (z) \end{pmatrix}=  \frac{1}{\sqrt{r_{u^{(n)} R}}}
\begin{pmatrix} \bar c_H S_R (z, \lambda_{u^{(n)}}, c_u) \cr
\noalign{\kern 5pt}
- i\bar s_H  \check C_R (z, \lambda_{u^{(n)}}, c_u) \end{pmatrix} ,
\label{wave-up1}
\end{align}
where $\check S_L$ and $\check C_R$ are given in (\ref{functionA2}) and 
$(\bar c_H, \bar s_H) = (\cos \onehalf \theta_H, \sin \onehalf \theta_H)$. 
The normalization factor is determined by the condition
\begin{align}
\int_1^{z_L} dz \, \Big\{ |f (z) |^2 + |g (z) |^2 \Big\} = 1 \quad {\rm for~} 
\begin{pmatrix} f (z) \cr g (z) \end{pmatrix} .
\label{normalizationF1}
\end{align}
One can show that $r_{u^{(n)} L} = r_{u^{(n)} R} $ for $\lambda_{u^{(n)}} \not= 0$.  % Similar formulas hold for charm and top towers.
Wave functions in the original gauge are given by 
\begin{align}
&\begin{pmatrix}  f^{u^{(n)}}_{L/R} (z) \cr g^{u^{(n)}}_{L/R}  (z) \end{pmatrix}
=\begin{pmatrix} \cos \onehalf \theta(z) & i \sin  \onehalf \theta(z) \cr
\noalign{\kern 3pt}
i \sin  \onehalf \theta(z) & \cos \onehalf \theta(z)  \end{pmatrix}
\begin{pmatrix} \tilde f^{u^{(n)}}_{L/R}  (z) \cr \tilde g^{u^{(n)}}_{L/R}  (z) \end{pmatrix} .
\label{wave-up2}
\end{align}

In the  down-type quark sector,  $d$, $d^{\prime}$, $D_d^{+ }$ and $D_d^{- }$ intertwine with each other in the first generation.
The KK expansion is given by
\begin{align}
 \begin{pmatrix}\check  d \cr \check d'  \cr \check D_d^+ \cr \check D_d^- \end{pmatrix} 
&=  \sqrt{k} \sum_{n=0}^\infty \Bigg\{ d^{(n)}_L (x) 
\begin{pmatrix} f^{d^{(n)}}_L (z) \cr g^{d^{(n)}}_L (z) \cr h^{d^{(n)}}_L (z) \cr k^{d^{(n)}}_L (z)\end{pmatrix}
+ d^{(n)}_R (x) 
\begin{pmatrix} f^{d^{(n)}}_R (z) \cr g^{d^{(n)}}_R (z)  \cr h^{d^{(n)}}_R (z) \cr k^{d^{(n)}}_R (z) \end{pmatrix} \Bigg\} \cr
\noalign{\kern 5pt}
&\hskip -.3cm 
+  \sqrt{k} \sum_{n=1}^\infty \Bigg\{ D^{(n)}_{d L} (x) 
\begin{pmatrix} f^{D_d^{(n)}}_L (z) \cr g^{D_d^{(n)}}_L (z) \cr h^{D_d^{(n)}}_L (z) \cr k^{D_d^{(n)}}_L (z)\end{pmatrix}
+ D^{(n)}_{dR} (x) 
\begin{pmatrix} f^{D_d^{(n)}}_R (z) \cr g^{D_d^{(n)}}_R (z)  \cr h^{D_d^{(n)}}_R (z) \cr k^{D_d^{(n)}}_R (z) \end{pmatrix} \Bigg\}
\label{wave-down1}
\end{align}
where the wave functions are normalized by 
\begin{align}
&\int_1^{z_L} dz   \, \big\{ | f |^2 + |g |^2 + |h|^2 + |k |^2 \big\} = 1 
\label{normalization-down1}
\end{align}
in each mode.
The lowest mode $d^{(0)}$ is the $d$ quark.  In the wave functions of the $d$ and $D$ towers,   $(f,g)$  and $(h,k)$
components are dominant, respectively.  The mass spectrum $\{ m_n= k \lambda_n \}$ is determined by 
\begin{align}
&\Big( S_L S_R +\sin^2\frac{\theta_H}{2} \Big)
 \big({\cal S}_{L1} {\cal S}_{R1}  -{\cal S}_{L2} {\cal S}_{R2} \big)   \cr
 \noalign{\kern 5pt}
 &\hskip 1.5cm
+|\mu_d  |^2 C_R S_R
 \big( {\cal S}_{L1} {\cal C}_{L1}  -{\cal S}_{L2}  {\cal C}_{L2}  \big)=0 ~, \cr
 \noalign{\kern 5pt}
&(C, S)_{L/R} =(C, S)_{L/R} (z; \lambda_n, c_u) \big|_{z=1}  , \cr
%\noalign{\kern 5pt}
%S_{L/R}(z) &=S_{L/R} (z; \lambda_{\alpha^{(n)}} , c_{q^\alpha}) , \cr
%\noalign{\kern 5pt}
&({\cal C, S})_{L/R j} = ({\cal C, S})_{L/R j}(z;\lambda_n,  c_{D_d}, \tilde m_{D_d}) \big|_{z=1} , 
%\noalign{\kern 5pt}
%{\cal S}_{L/R j}(z) &= {\cal S}_{L/R j}(z;\lambda_{\alpha^{(n)}}, c_{D^\alpha}, \tilde m_\alpha) , 
\label{downSpectrum1}
\end{align}
where $({\cal C, S})_{L/R j}(z;\lambda,  c_, \tilde m)$  is given in Eq.\ (\ref{MassiveFermion1}).
$\mu_d$ arises from a brane interaction.  $c_{D_d}$ is the bulk mass parameter of $D_d^\pm$ fields.
$k \tilde m_{D_d}$ is a vector-like mass connecting $D_d^+$ and $D_d^-$ fields.
Details of the wave functions are found in Ref.\ \cite{YHbook}.
The wave functions $(f^{d^{(n)}}_{L /R} , g^{d^{(n)}}_{L/R} )$ and $(f^{D_d^{(n)}}_{L /R} , g^{D_d^{(n)}}_{L/R} )$ 
in the original gauge are related to 
$(\tilde f^{d^{(n)}}_{L /R} , \tilde g^{d^{(n)}}_{L/R} )$ and $(\tilde f^{D_d^{(n)}}_{L /R} , \tilde g^{D_d^{(n)}}_{L/R} )$ 
in the twisted gauge by the same formula as in Eq.\ (\ref{wave-up2}).

The spectrum $\{  m_{e^{(n)}} = k  \lambda_{e^{(n)}} \}$ of the electron KK tower is determined by 
\begin{align}
&S_L (1;  \lambda_{e^{(n)}}, c_e)  S_R (1;  \lambda_{e^{(n)}}, c_e) +\sin^2\onehalf \theta_H =0 ~.
%&S_L (1, \lambda, c_q)  S_R (1, \lambda, c_q) +\sin^2\frac{\theta_H}{2}=0 ~.
\label{electron-mass1}
\end{align}
The electron mass $m_e = m_{e^{(0)}} $ fixes $c_e$.  The KK expansion is given by
\begin{align}
& \begin{pmatrix} \tilde{\check e} \cr \tilde {\check e}' \end{pmatrix} = 
\sqrt{k} \sum_{n=0}^\infty \bigg\{ e^{(n)}_L (x) \begin{pmatrix} f^{e^{(n)}}_L (z) \cr g^{e^{(n)}}_L (z) \end{pmatrix}
+e^{(n)}_R (x) \begin{pmatrix} f^{e^{(n)}}_R (z) \cr g^{e^{(n)}}_R (z) \end{pmatrix} \bigg\} .
\label{electronKK1}
\end{align}
Wave functions take the same form as those in Eq.\ (\ref{wave-up1}) where $u^{(n)}$ and $c_u$ are
replaced by $e^{(n)}$ and $c_e$.

In the neutrino sector $\nu_e$, $\nu_e^{\prime}$ and $\hat \chi^1 = (\eta_e^c , \eta_e )$ fields intertwine with each other.
The spectrum of the $\nu_e$ KK tower, 
$\{ \nu_e^{+ (n)},   \nu_e^{- (\ell)} ; n \ge 0, \ell \ge 1 \}$, is determined by
\begin{align}
\det K_{\nu_e^\pm} = 
(k  \lambda_{\nu_e^{\pm (n)}} \mp  M_e )
\Big\{ S_L^e S_R^e +\sin^2\frac{\theta_H}{2} \Big\}
+ \frac{|m_{B_e}|^2}{2 \, k} S_R^e C_R^e =0 ~.
\label{neutrinoSpectrum1}
\end{align}
Here $S_{L/R}^e = S_{L/R} (1; \lambda_{\nu_e^{\pm (n)}}, c_e)$ etc.  $M_e$ is a Majorana mass of $\hat \chi^1$ and
$m_{B_e}$ arises from a brane interaction.
$\nu_e^{+ (0)}$ is the light neutrino field. Its tiny mass generated  by a gauge-Higgs seesaw mechanism 
is $m_{\nu_e} \sim m_e^2 M_e / (|c_e|- \onehalf) m_{B_e}^2$.
The KK expansion is given by
\begin{align}
\begin{pmatrix} \check \nu_e \cr \check \nu_e' \end{pmatrix}  &=  
\sqrt{k} \sum_{n=0}^\infty \Bigg\{ \nu^{+(n)}_{eL} (x) \begin{pmatrix} f_L^{\nu_e^{+(n)}} (z) \cr g_L^{\nu_e^{+(n)}} (z) \end{pmatrix}
+ \nu^{+(n)}_{eR} (x) \begin{pmatrix} f_R^{\nu_e^{+(n)}} (z) \cr g_R^{\nu_e^{+(n)}} (z) \end{pmatrix} \Bigg\} \cr
\noalign{\kern 5pt}
%&\hskip 1.5cm 
&+ \sqrt{k} \sum_{n=1}^\infty  \Bigg\{ \nu^{-(n)}_{e L} (x) \begin{pmatrix} f_L^{\nu_e^{-(n)}} (z) \cr g_L^{\nu_e^{-(n)}} (z) \end{pmatrix}
+ \nu^{-(n)}_{e R} (x) \begin{pmatrix} f_R^{\nu_e^{-(n)}} (z) \cr g_R^{\nu_e^{-(n)}} (z) \end{pmatrix} \Bigg\} , \cr
\noalign{\kern 5pt}
%&\hskip .5cm
\eta_e &= \sum_{n=0}^\infty  \nu^{+(n)}_{e L} (x)  \, h^{\nu_e^{+(n)}} +  \sum_{n=1}^\infty  \nu^{-(n)}_{e L} (x)  \, h^{\nu_e^{-(n)}}  ~, \cr
\noalign{\kern 5pt}
&\quad
\nu^{\pm (n)}_{eR} = \pm  ( \nu^{\pm (n)}_{eL} )^c = \pm e^{i\delta_C} \sigma^2 ( \nu^{\pm (n)}_{eL} )^* ~.
\label{wave-neutrino1}
\end{align}
In each mode $\int_1^{z_L} dz   \, \big\{ | f_L |^2 + |g_L |^2 + | f_R |^2 + |g_R |^2 \big\}  + |h |^2 = 1$.
Details of wave functions have been given in Ref.\ \cite{YHbook}.

\section{Gauge couplings} 

Gauge couplings of fermion fields are contained in the part of the action 
\begin{align}
&\int d^4 x \int_1^{z_L} \frac{dz}{k} \sum_J \bar{\check \Psi}^J \gamma^\mu (-i)
\Big(  g_A A_\mu + g_B Q_X B_\mu + g_C A_\mu^{SU(3)_C}  \Big)\check  \Psi^J 
\label{GHUgaugecoupling1}
\end{align}
where $\check \Psi^J = z^{-2} \Psi^J $.  
Inserting the KK expansions for $A_\mu$, $B_\mu$, $A_\mu^{SU(3)_C}$, $\check \Psi^J$ and $\bar{\check \Psi}^J$,
obtained in the twisted gauge,  
into (\ref{GHUgaugecoupling1}) and integrating over $z$, one finds gauge couplings of all
fermion modes.

The $W$ interaction is evaluated from % in the twisted gauge,  from
\begin{align}
{\cal L}^W_{\rm int} &= - ig_A \sum_J  \int_1^{z_L} \frac{dz}{\sqrt{k}} \, 
 \overline{\tilde{\check \Psi}}{}^J
\gamma^\mu
\sum_{a=1}^2 ( \tilde A_\mu^{a_L}  T^{a_L} + \tilde A_\mu^{a_R}  T^{a_R} + \tilde  A_\mu^{\hat a} T^{\hat a} )
\tilde{\check \Psi}{}^J ~.
\label{Wfermion1}
\end{align}
%where $\sum_J$ extends over $\Psi^J = \Psi^q, \Psi^\ell, \Psi^{F_q}$ and $\Psi^{F_\ell}$.
The $W ud$ couplings are given by
\begin{align}
%&{\cal L}^{W ud} = 
{\cal L}^{W ud} &= -i \frac{g_w}{\sqrt{2}} \sum_{\ell=0}^\infty  W_\mu^{(\ell) \dagger} 
 \sum_{n, m =0}^\infty  \Big\{ 
\hat g^{W ud}_{L\, \ell n m} \bar u_L^{(n)} \gamma^\mu d_L^{(m)}
+ \hat g^{W ud}_{R\, \ell n m} \bar u_R^{(n)} \gamma^\mu d_R^{(m)} \Big\} + {\rm H.c.} ~, \cr
\noalign{\kern 5pt}
\hat g^{W ud}_{L/R\, \ell n m} &=  G_W[ (\tilde h^L, \tilde  h^R, \tilde {\hat h})_{W^{(\ell)}}; 
( \tilde f,\tilde g)^{u^{(n)}}_{L/R} ,   (\tilde f, \tilde g)^{d^{(m)}}_{L/R} ] \cr
\noalign{\kern 5pt}
&
= G_W[ (h^L, h^R, {\hat h})_{W^{(\ell)}}; 
(  f, g)^{u^{(n)}}_{L/R} ,   (  f,  g)^{d^{(m)}}_{L/R} ] ,  
\label{Wud1}
\end{align}
where 
\begin{align}
&G_W [(h^L, h^R, \hat h)_\alpha; (f, g)_1, (f, g)_2]  \cr
\noalign{\kern 2pt}
&%\quad
= \sqrt{kL} \int_1^{z_L} dz \, \Big\{ h^{L*}_\alpha   f_1^*   f_2 +  h^{R*}_\alpha  g_1^*   g_2
+  \frac{i}{\sqrt{2}} \,  \hat h^*_\alpha  ( f_1^*  g_2 -  g_1^*  f_2 ) \Big\} ~.
\label{GWfunction1}
\end{align}
The last equality in Eq.\ (\ref{Wud1}) is confirmed with the aid of Eqs.\ (\ref{WbosonWave2}) and (\ref{wave-up2}).
The couplings to the $D^{(m)}$ ($m \ge 1$) modes are given by replacing $d^{(m)}$ by  $D^{(m)}$
in the above formula.
The $W \nu_e e$ couplings are given by
\begin{align}
{\cal L}^{W \nu_e e} &= -i \frac{g_w}{\sqrt{2}} \sum_{\ell =0}^\infty  W_\mu^{(\ell) \dagger}  %\cr
%\noalign{\kern 5pt}
% &\quad
%\times
 \bigg[ \sum_{n,m=0}^\infty 
\Big\{ \hat g^{W\nu^+_{e} e}_{L\, \ell nm} \,  \bar \nu_{eL}^{+(n)} \gamma^\mu e_L^{(m)} 
+ \hat g^{W\nu^+_{e} e}_{R\, \ell nm} \,   \bar \nu_{eR}^{+(n)} \gamma^\mu e_R^{(m)} \Big\} \cr
\noalign{\kern 5pt}
&\qquad
+  \sum_{n=1}^\infty   \sum_{m=0}^\infty 
\Big\{ \hat g^{W\nu^-_{e} e}_{L\, \ell nm} \,  \bar \nu_{eL}^{-(n)} \gamma^\mu e_L^{(m)} 
+ \hat g^{W\nu^-_{e} e}_{R\, \ell nm} \,   \bar \nu_{eR}^{-(n)} \gamma^\mu e_R^{(m)} \Big\} \bigg] + {\rm H.c.} ~, \cr
\noalign{\kern 5pt}
\hat g^{W\nu^\pm_{e} e}_{L/R \, \ell nm} &=
 G_W[ (\tilde h^L, \tilde h^R, \tilde {\hat h})_{W^{(\ell)}}; 
( \tilde f, \tilde g)^{\nu_e^{\pm (n)}}_{L/R} ,  (\tilde f, \tilde g)^{e^{(m)}}_{L/R}]  \cr
\noalign{\kern 5pt}
&= G_W[ ( h^L,  h^R,  {\hat h})_{W^{(\ell)}}; 
(  f, g)^{\nu_e^{\pm (n)}}_{L/R} ,  (f, \ g)^{e^{(m)}}_{L/R}] ~.
\label{Wnue1}
\end{align}

Similarly the $Z$ interaction is evaluated from
\begin{align}
{\cal L}_{\rm int}^{Z} &= - ig_A \sum_J  \int_1^{z_L} \frac{dz}{\sqrt{k}} 
 \overline{\tilde{\check \Psi}} {}^J \gamma^\mu
\Big(  \tilde A_\mu^{3_L}  T^{3_L} + \tilde A_\mu^{3_R}  T^{3_R}   +  \tilde A_\mu^{\hat 3}  T^{\hat 3} 
+ \frac{g_B}{g_A}   B_\mu Q_X
 \Big) \tilde{\check  \Psi}^J .
\label{Zfermion1}
\end{align}
The $Z uu$ and $Z dd$ couplings are given by
\begin{align}
&{\cal L}^{Z ,u, d}  =-i  \frac{g_w}{\cos \theta_W^0}  \sum_{\ell=0}^\infty  Z_\mu^{(\ell)} 
%\cr \noalign{\kern 3pt} &\qquad  \times 
 \sum_{n, m =0}^\infty  \Big\{ 
\hat g^{Z uu}_{L\, \ell n m} \bar u_L^{(n)} \gamma^\mu u_L^{(m)}
+ \hat g^{Z uu}_{R\, \ell n m} \bar u_R^{(n)} \gamma^\mu u_R^{(m)} \cr
\noalign{\kern 3pt}
&\hskip 4.cm
+ \hat g^{Zdd}_{L\, \ell n m} \bar d_L^{(n)} \gamma^\mu d_L^{(m)}
+ \hat g^{Z dd}_{R\, \ell n m} \bar d_R^{(n)} \gamma^\mu d_R^{(m)} \Big\} ,  
\label{Zud1}
\end{align}
where
\begin{align}
\begin{pmatrix} \hat g^{Z uu}_{L/R\, \ell n m} \cr \mynoalign \hat g^{Z dd}_{L/R\, \ell n m} \end{pmatrix}
&= \begin{pmatrix} \hat g^{Z uu, su2}_{L/R\, \ell n m}  \cr \mytinynoalign \hat g^{Z dd, su2}_{L/R\, \ell n m} \end{pmatrix}
- \sin^2 \theta_W^0 \,  \begin{pmatrix} \hat g^{Z uu, em}_{L/R\, \ell n m} \cr 
 \mytinynoalign \hat g^{Z dd, em}_{L/R\, \ell n m} \end{pmatrix} ,  \cr
\noalign{\kern 5pt}
\begin{pmatrix} \hat g^{Z uu, su2}_{L/R\, \ell n m}  \cr \mytinynoalign \hat g^{Z dd, su2}_{L/R\, \ell n m} \end{pmatrix}
&= \cos \theta_W^0  \, 
\begin{pmatrix} 
T^3_u \, G_W[ (\tilde h^L, \tilde h^R, \tilde {\hat h})^{su2}_{Z^{(\ell)}};  (\tilde f, \tilde g)^{u^{(n)}}_{L/R} , 
(\tilde f, \tilde g)^{u^{(m)}}_{L/R} ]  \cr
T^3_d \, G_W[ (\tilde h^L, \tilde h^R, \tilde {\hat h})^{su2}_{Z^{(\ell)}};  (\tilde f, \tilde g)^{d^{(n)}}_{L/R} , 
(\tilde f, \tilde g)^{d^{(m)}}_{L/R} ]  
\end{pmatrix} , \cr
\noalign{\kern 5pt}
& = \cos \theta_W^0 
\begin{pmatrix}
T^3_u \, G_W[ (h^L, h^R, \hat h)^{su2}_{Z^{(\ell)}};  (f, g)^{u^{(n)}}_{L/R} , (f, g)^{u^{(m)}}_{L/R} ] \cr
T^3_d \, G_W[ (h^L, h^R, \hat h)^{su2}_{Z^{(\ell)}};  (f, g)^{d^{(n)}}_{L/R} , (f, g)^{d^{(m)}}_{L/R} ] 
\end{pmatrix} , \cr
\noalign{\kern 5pt}
 \begin{pmatrix} \hat g^{Z uu, em}_{L/R\, \ell n m} \cr 
 \mytinynoalign \hat g^{Z dd, em}_{L/R\, \ell n m} \end{pmatrix}
&=  \cos \theta_W^0 
\begin{pmatrix}
 Q_u \, G_\gamma^u [ h^{em}_{Z^{(\ell)}} ;   (\tilde f, \tilde g)^{u^{(n)}}_{L/R},   (\tilde f, \tilde g)^{u^{(m)}}_{L/R} ] \cr
 Q_d \, G_\gamma^d [ h^{em}_{Z^{(\ell)}} ;  (\tilde f, \tilde g, h, k)^{d^{(n)}}_{L/R}, (\tilde f,\tilde g, h, k)^{d^{(m)}}_{L/R}  ] 
 \end{pmatrix} , \cr
\noalign{\kern 5pt}
&=  \cos \theta_W^0 
\begin{pmatrix}
 Q_u \, G_\gamma^u [ h^{em}_{Z^{(\ell)}} ;   (f, g)^{u^{(n)}}_{L/R},   (f, g)^{u^{(m)}}_{L/R} ] \cr
 Q_d \, G_\gamma^d [ h^{em}_{Z^{(\ell)}} ;  (f, g, h, k)^{d^{(n)}}_{L/R}, (f,g, h, k)^{d^{(m)}}_{L/R}  ] 
 \end{pmatrix} ,
\label{Zud2}
\end{align}
$(T^3_u, T^3_d) = (\onehalf, - \onehalf)$ and $(Q_u, Q_d) = (\frac{2}{3}, - \frac{1}{3} )$.
Here 
\begin{align}
&G_\gamma^u [h_\gamma ; (f, g)_1 , (f, g)_2 ]  
= \sqrt{kL} \int_1^{z_L} dz  \, h_\gamma  \big( f_1^*  f_2 + g_1^* g_2  \big) , \cr
\noalign{\kern 5pt}
&G_\gamma^d [h_\gamma ; (f, g, h, k)_1 ,  (f, g, h, k)_2 ]  \cr
\noalign{\kern 3pt}
&\hskip 1.cm
= \sqrt{kL} \int_1^{z_L} dz  \, h_\gamma  \big( f_1^*  f_2 + g_1^* g_2 + h_1^*  h_2 + k_1^* k_2 \big) .
\label{Ggammafunction1}
\end{align}
The $Z^{(0)}$ couplings to $d^{(n)} D_d^{(m)}$ and $D_d^{(n)} D_d^{(m)}$ are very small (
$|\hat g^{ZdD_d}_{L/R \, 0nm}| , |\hat g^{ZD_dD_d}_{L/R \, 0nm}| < O(10^{-6})$) 
except for diagonal elements
$\hat g^{ZD_dD_d}_{L/R \,  0 nn}$ $(n \ge 1)$ for which there are contributions from the $U(1)$ part.
Numerically $\hat g^{ZD_dD_d, em}_{L/R \,  0 nn} \sim - 0.333385$ and 
$|\hat g^{ZD_dD_d, su2}_{L/R \,  0 nn}| < O(10^{-12})$ for $\theta_H=0.1$ and $m_\KK = 13\,$TeV.

In a similar manner the $Zee$ and $Z\nu_e \nu_e$ couplings are found to be
\begin{align}
&\hat g^{Z ee}_{L/R\, \ell n m} = \hat g^{Z ee, su2}_{L/R\, \ell n m} - \sin^2 \theta_W^0 \, \hat g^{Z ee, em}_{L/R\, \ell n m} ~, \cr
\noalign{\kern 3pt}
&\hat g^{Zee, su2}_{L/R\, \ell n m} = \cos \theta_W^0 T^3_e \, 
G_W[ (h^L, h^R, \hat h)^{su2}_{Z^{(\ell)}};  (f, g)^{e^{(n)}}_{L/R} , (f, g)^{e^{(m)}}_{L/R} ] , \cr
\noalign{\kern 3pt}
&\hat g^{Z ee, em}_{L/R\, \ell n m} =  \cos \theta_W^0 Q_e \,
G_\gamma^u [ h^{em}_{Z^{(\ell)}} ;   (f, g)^{e^{(n)}}_{L/R},   (f, g)^{e^{(m)}}_{L/R} ] , \cr
\noalign{\kern 5pt}
&\hat g^{Z \nu_e^a \nu_e^b}_{L/R\, \ell n m} = \hat g^{Z \nu_e^a \nu_e^b, su2}_{L/R\, \ell n m}  \cr
\noalign{\kern 3pt}
&\hskip 0.6cm
= \cos \theta_W^0 T^3_{\nu}\, 
G_W[ (h^L, h^R, \hat h)^{su2}_{Z^{(\ell)}};  (f, g)^{\nu_e^{a (n)}}_{L/R} , (f, g)^{\nu_e^{b(m)}}_{L/R} ] ,
\label{Zenu2}
\end{align}
where $a,b=+ ~{\rm or}~ -$, $Q_e=-1$ and  $(T^3_{\nu}, T^3_e) = (\onehalf, - \onehalf)$.
Here $\hat g^{Z \nu_e^- \nu_e^\pm}_{L/R\, \ell 0 m} = \hat g^{Z \nu_e^\pm \nu_e^-}_{L/R\, \ell n 0} = 0$ has been understood.

\begin{figure}[tbh]
\centering
\includegraphics[height=50mm]{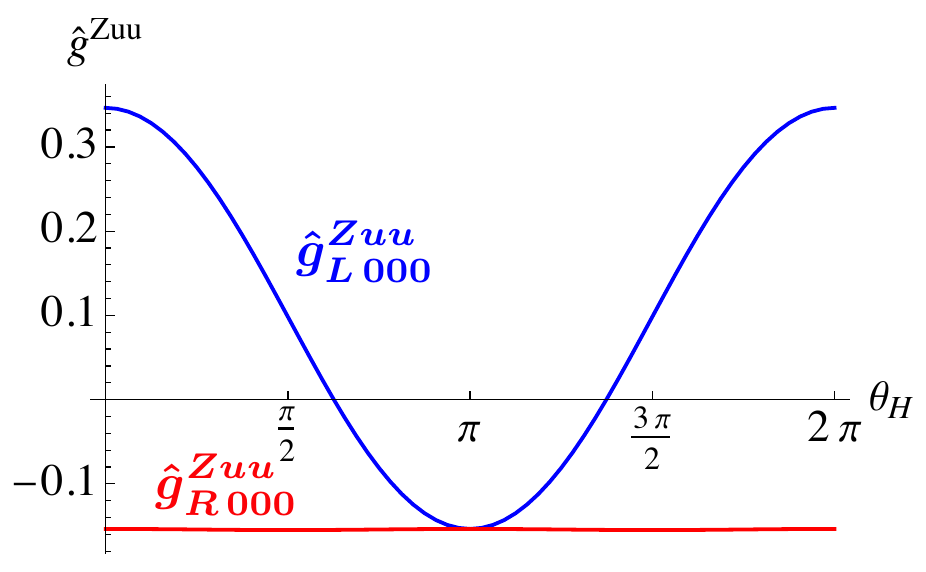}
\caption{The $\theta_H$-dependence of $Z$-boson couplings of $u$ quarks,  $\hat g^{Zuu}_{L \, 000}$ and $\hat g^{Zuu}_{R \, 000}$, 
is displayed.  $\hat g^{Zuu}_{R \, 000} (\theta_H)$ is nearly constant; $- 0.15495 < \hat g^{Zuu}_{R \, 000} < -0.15357$.
}
\label{fig:gZuu}
\end{figure}

We note that all gauge couplings depend on $\theta_H$.  The $Z= Z^{(0)}$ coupling of up quarks $u^{(0)}$, 
$\hat g^{Zuu}_{L/R \, 000} $, for instance, 
takes the SM value $\hat g^{Zuu}_{L \, 000} = T^3_u - \sin^2 \theta_W^0 Q_u$ and 
$\hat g^{Zuu}_{R \, 000}  =  - \sin^2 \theta_W^0 Q_u$ at $\theta_H=0$.    
However it varies with $\theta_H$.  
The $\theta_H$-dependence of $\hat g^{Zuu}_{L/R \, 000}$  is depicted in Fig.\  \ref{fig:gZuu}.
The couplings become vector-like,  $\hat g^{Zuu}_{L \, 000} = \hat g^{Zuu}_{R \, 000}$, at $\theta_H = \pi$,
where the $Z$ boson is massless and the $u$ quark is massive.
The $\theta_H$-dependence of $m_Z$ and $m_u$ is qualitatively similar to that of $m_W$ and $m_t$
in Fig.\ \ref{fig:mWtop}.

%%%%%%%%%%%%%%%%%%%%
%%%%%%%%%%%%%%%%%%%

\section{Chiral anomalies}

\begin{figure}[tbh]
\centering
\includegraphics[height=24mm]{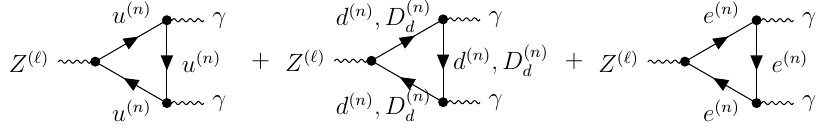}
\caption{Chiral anomalies for $\gamma \gamma  Z^{(\ell)}$ in GHU.
}
\label{fig:gamgamZ}
\end{figure}

\begin{figure}[tbh]
\centering
\includegraphics[height=25mm]{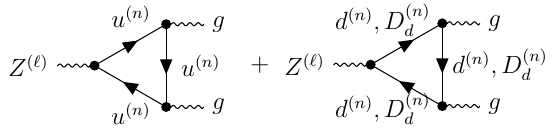}
\caption{Chiral anomalies for $g g  Z^{(\ell)}$ in GHU.
}
\label{fig:glueglueZ}
\end{figure}

Chiral anomalies in four dimensions are expressed in terms of gauge couplings obtained in the previous section. 
Gauge anomalies for $\gamma \gamma Z^{(\ell)}$ depicted in Fig.\  \ref{fig:gamgamZ} are proportional to 
\begin{align}
J_{\gamma \gamma Z^{(\ell)}} = &3 \Big\{ Q_u^2  (\Tr \hat g^{Zuu}_{L \, \ell} - \Tr \hat g^{Zuu}_{R \, \ell} )
+ Q_d^2  (\Tr \hat g^{Zdd}_{L \, \ell} - \Tr \hat g^{Zdd}_{R \, \ell} )  \cr
\noalign{\kern 5pt}
&%\hskip 1.cm
+ Q_d^2  (\Tr \hat g^{ZD_d D_d}_{L \, \ell} - \Tr \hat g^{ZD_d D_d}_{R \, \ell} )  \Big\} 
+ Q_e^2  (\Tr \hat g^{Zee}_{L \, \ell} - \Tr \hat g^{Zee}_{R \, \ell} ) ~, \cr
\noalign{\kern 5pt}
& ( \hat g^{Zuu}_{L/R \, \ell} )_{nm} = \hat g^{Zuu}_{L/R \, \ell nm} ~{\rm etc.} 
\label{anomalyJ1}
\end{align}
Note that KK excited modes of quarks and leptons contribute to anomalies.
In the down quark sector both $\{ d^{(n)} \}$ and $\{ D_d^{(n)} \}$ towers contribute.
Similarly  for $ggZ^{(\ell)}$ (where $g$ is a gluon) anomalies depicted in Fig.\  \ref{fig:glueglueZ}  are proportional to
\begin{align}
J_{gg Z^{(\ell)}} = &\Tr \hat g^{Zuu}_{L \, \ell} - \Tr \hat g^{Zuu}_{R \, \ell} 
+  \Tr \hat g^{Zdd}_{L \, \ell} - \Tr \hat g^{Zdd}_{R \, \ell}   \cr
\noalign{\kern 5pt}
&\hskip .5cm
+  \Tr \hat g^{ZD_d D_d}_{L \, \ell} - \Tr \hat g^{ZD_d D_d}_{R \, \ell}  ~.
\label{anomalyJ2}
\end{align}
Cancellation of these two anomalies is achieved when
\begin{align}
&\Tr \hat g^{Zuu}_{L \, \ell} - \Tr \hat g^{Zuu}_{R \, \ell}  \cr
&= -  ( \Tr \hat g^{Zdd}_{L \, \ell} - \Tr \hat g^{Zdd}_{R \, \ell}  + \Tr \hat g^{ZD_d D_d}_{L \, \ell} - \Tr \hat g^{ZD_d D_d}_{R \, \ell}  ) \cr
&= - ( \Tr \hat g^{Zee}_{L \, \ell} - \Tr \hat g^{Zee}_{R \, \ell} ) ~.
\label{cancell1}
\end{align}

%%%%%%%%%%%%%%%%%%%%
%%%%%%%%%%%%%%%%%%%

%%%%%%%%%%%%%%%%%%%%
%%%%%%%%%%%%%%%%%%%

\begin{figure}[tbh]
\centering
\includegraphics[height=24mm]{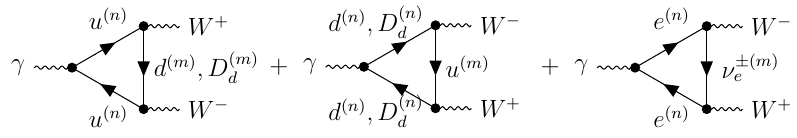}
\caption{Chiral anomalies for $\gamma WW$ in GHU.
}
\label{fig:gammaWW}
\end{figure}

\begin{figure}[tbh]
\centering
\includegraphics[height=25mm]{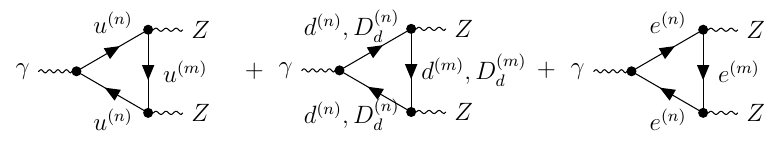}
\caption{Chiral anomalies for $\gamma ZZ$ in GHU.
}
\label{fig:gammaZZ}
\end{figure}

Anomalies for the vertex $\gamma W W$ depicted in Fig.\ \ref{fig:gammaWW} are proportional to
\begin{align}
J_{\gamma WW} & = 3 \,  Q_u \sum_{\tilde d = d, D_d}
( \Tr \hat g^{Wu \tilde d}_{L \, 0}  \hat g^{W^\dagger \tilde d u}_{L \, 0}  
- \Tr \hat g^{Wu \tilde d}_{R \, 0}  \hat g^{W^\dagger \tilde d u}_{R \, 0}  ) \cr
\noalign{\kern 5pt}
&+ 3 \, Q_d  \sum_{\tilde d = d, D_d}
( \Tr \hat g^{W^\dagger \tilde du}_{L \, 0}   \hat g^{Wu \tilde d}_{L \, 0} 
- \Tr \hat g^{W^\dagger \tilde du}_{R \, 0}   \hat g^{Wu \tilde d}_{R \, 0} ) \cr
\noalign{\kern 5pt}
&%\hskip 1.cm
+ Q_e \sum_{a=+,-}  (\Tr \hat g^{W^\dagger e \nu_e^a}_{L \, 0}   \hat g^{W\nu_e^a e}_{L \, 0} 
- \Tr \hat g^{W^\dagger e \nu_e^a}_{R\, 0}   \hat g^{W\nu_e^a e}_{R \, 0}  ) ~, \cr
\noalign{\kern 5pt}
&\quad
 \hat g^{W^\dagger du}_{L/R \, \ell} = (\hat g^{W ud}_{L/R \, \ell} )^\dagger ~ {\rm etc.} 
%& ( \hat g^{W^\dagger du}_{L/R \, 0} )_{nm} = (\hat g^{W ud}_{L/R \, 0 mn} )^* ~{\rm etc.} 
\label{anomalyJ3}
\end{align}
Anomalies for the vertex $\gamma ZZ$ depicted in Fig.\ \ref{fig:gammaZZ} are proportional to
\begin{align}
J_{\gamma ZZ} &= 3 \, Q_u ( \Tr \hat g^{Zuu}_{L \, 0}  \hat g^{Zuu}_{L \, 0}  
- \Tr \hat g^{Zuu}_{R \, 0}  \hat g^{Zuu}_{R \, 0} ) \cr
\noalign{\kern 5pt}
&+ 3 \, Q_d  \sum_{\tilde d_1 , \tilde d_2 = d, D_d} 
( \Tr \hat g^{Z \tilde d_1 \tilde d_2}_{L \, 0}   \hat g^{Z \tilde d_2 \tilde d_1}_{L \, 0} 
- \Tr \hat g^{Z \tilde d_1 \tilde d_2}_{R \, 0}   \hat g^{Z \tilde d_2 \tilde d_1}_{R \, 0}  ) \cr
\noalign{\kern 5pt}
&%\hskip 1.cm
+ Q_e (\Tr \hat g^{Zee}_{L \, 0}   \hat g^{Zee}_{L \, 0} 
- \Tr \hat g^{Zee}_{R\, 0}   \hat g^{Zee}_{R \, 0}  ) ~.
%& ( \hat g^{W^\dagger du}_{L/R \, 0} )_{nm} = (\hat g^{W ud}_{L/R \, 0 mn} )^* ~{\rm etc.} 
\label{anomalyJ4}
\end{align}

%%%%%%%%%%%%%%%%%%%%
%%%%%%%%%%%%%%%%%%%

%%%%%%%%%%%%%%%%%%%%
%%%%%%%%%%%%%%%%%%%

\begin{figure}[tbh]
\centering
\includegraphics[height=47mm]{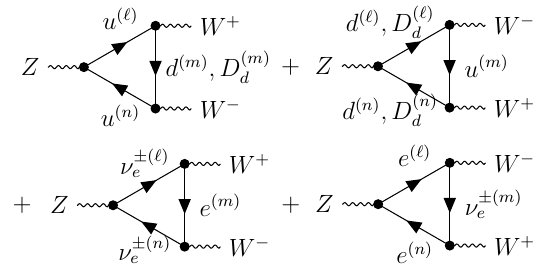}
\caption{Chiral anomalies for $ZWW$ in GHU.
}
\label{fig:ZWW}
\end{figure}

\begin{figure}[tbh]
\centering
\includegraphics[height=47mm]{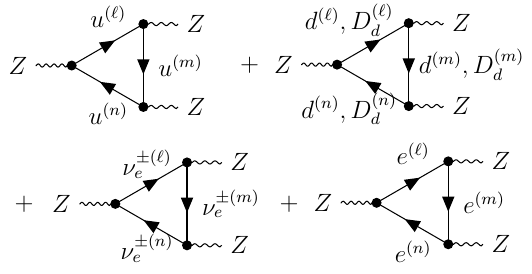}
\caption{Chiral anomalies for $ZZZ$ in GHU.
}
\label{fig:ZZZ}
\end{figure}

Anomalies for the vertex $ZWW$ depicted in Fig.\ \ref{fig:ZWW} are proportional to
\begin{align}
J_{ZWW} & = 3  \sum_{\tilde d = d, D_d}
( \Tr \hat g^{Wu \tilde d}_{L \, 0}  \hat g^{W^\dagger \tilde d u}_{L \, 0}  \hat g^{Zuu}_{L \, 0} 
- \Tr \hat g^{Wu \tilde d}_{R \, 0}  \hat g^{W^\dagger \tilde d u}_{R \, 0}   \hat g^{Zuu}_{R \, 0}  ) \cr
\noalign{\kern 5pt}
&+ 3   \sum_{\tilde d_1, \tilde d_2 = d, D_d}
( \Tr \hat g^{W^\dagger \tilde d_1 u}_{L \, 0}   \hat g^{Wu \tilde d_2}_{L \, 0}   \hat g^{Z \tilde d_2 \tilde d_1}_{L \, 0} 
- \Tr \hat g^{W^\dagger \tilde d_1 u}_{R \, 0}   \hat g^{Wu \tilde d_2}_{R \, 0}  \hat g^{Z \tilde d_2 \tilde d_1}_{R \, 0} ) \cr
\noalign{\kern 5pt}
&%\hskip 1.cm
+  \sum_{a, b=+,-}  (\Tr \hat g^{W  \nu_e^a e}_{L \, 0}   \hat g^{W^\dagger e \nu_e^b}_{L \, 0} \hat g^{Z  \nu_e^b  \nu_e^a}_{L \, 0} 
- \Tr  \hat g^{W\nu_e^a e}_{R \, 0}   \hat g^{W^\dagger e \nu_e^b}_{R\, 0} \hat g^{Z  \nu_e^b  \nu_e^a}_{R \, 0}  )  \cr
\noalign{\kern 5pt}
&%\hskip 1.cm
+  \sum_{a = +,-}  (\Tr \hat g^{W^\dagger e \nu_e^a}_{L \, 0}   \hat g^{W\nu_e^a e}_{L \, 0}   \hat g^{Zee}_{L \, 0}
- \Tr \hat g^{W^\dagger e \nu_e^a}_{R\, 0}   \hat g^{W\nu_e^a e}_{R \, 0}   \hat g^{Zee}_{R \, 0} ) ~.
\label{anomalyJ5}
\end{align}
Finally anomalies for the vertex $ZZZ$ depicted in Fig.\ \ref{fig:ZZZ} are proportional to
\begin{align}
J_{ZZZ} & = 3   \,
( \Tr \hat g^{Zuu}_{L \, 0}  \hat g^{Zuu}_{L \, 0}  \hat g^{Zuu}_{L \, 0} 
- \Tr \hat g^{Zuu}_{R \, 0}  \hat g^{Zuu}_{R \, 0}   \hat g^{Zuu}_{R \, 0}  ) \cr
\noalign{\kern 5pt}
&+ 3   \sum_{\tilde d_1, \tilde d_2 , \tilde d_3 = d, D_d}
( \Tr \hat g^{Z \tilde d_1 \tilde d_2}_{L \, 0}   \hat g^{Z \tilde d_2 \tilde d_3}_{L \, 0}   \hat g^{Z \tilde d_3 \tilde d_1}_{L \, 0} 
- \Tr \hat g^{Z \tilde d_1 \tilde d_2}_{R \, 0}   \hat g^{Z \tilde d_2 \tilde d_3}_{R \, 0}   \hat g^{Z \tilde d_3 \tilde d_1}_{R \, 0}  ) \cr
\noalign{\kern 5pt}
&%\hskip 1.cm
+  \sum_{a, b, c =+,-}  
(\Tr \hat g^{Z  \nu_e^a \nu_e^b}_{L \, 0}   \hat g^{Z \nu_e^b \nu_e^c}_{L \, 0} \hat g^{Z  \nu_e^c  \nu_e^a}_{L \, 0} 
-  \Tr \hat g^{Z  \nu_e^a \nu_e^b}_{R \, 0}   \hat g^{Z \nu_e^b \nu_e^c}_{R \, 0} \hat g^{Z  \nu_e^c  \nu_e^a}_{R \, 0} ) \cr
\noalign{\kern 5pt}
&%\hskip 1.cm
+  (\Tr \hat g^{Zee}_{L \, 0}   \hat g^{Zee}_{L \, 0}   \hat g^{Zee}_{L \, 0}
- \Tr \hat g^{Zee}_{R\, 0}   \hat g^{Zee}_{R \, 0}   \hat g^{Zee}_{R \, 0} ) ~.
\label{anomalyJ6}
\end{align}

%%%%%%%%%%%%%%%%%%%%
%%%%%%%%%%%%%%%%%%%

%%%%%%%%%%%%%%%%%%%%
%%%%%%%%%%%%%%%%%%%

We show below that all of $J_{\gamma \gamma Z^{(\ell)}}$, $J_{ggZ^{(\ell)}}$, $J_{\gamma WW}$, 
$J_{\gamma ZZ}$, $J_{Z WW}$ and $J_{Z ZZ}$  vanish.  It is necessary to incorporate contributions of
KK fermions running inside loops in triangular diagrams.

\section{Holography in anomaly} 

In GHU in the RS warped space gauge couplings of quarks and leptons in four dimensions sightly
deviate from those in the SM.  They depend on particle species.  Although deviations are very tiny, 
cancellation of gauge anomalies would be spoiled if only contributions of quarks and leptons are
taken into account.  In GHU there are KK excitation modes of quarks and leptons, which also
give non-vanishing contributions to anomalies as summarized in the previous section.
In the RS space, or more generally in curved space, gauge couplings are not diagonal in 
the KK levels, and evaluation of total anomalies becomes highly nontrivial. 

It was shown in Ref.\ \cite{AnomalyFlow2}  that in an $SU(2)$ GHU model in the RS space 
total anomaly is expressed in terms of the values of wave functions of gauge fields at the UV brane (at $y=0$) and
IR brane (at $y=L$) and orbifold boundary conditions satisfied by fermion multiples.  It does not
depend on the fermion bulk mass parameter.  $\theta_H$ dependence is also described by
wave functions of gauge fields.  
In this paper we generalize the argument of Ref.\ \cite{AnomalyFlow2} to the GUT-inspired GHU model,
and derive holographic formulas for  anomalies.  Anomalies are expressed in terms of the values of
$W$ and $Z$ wave functions at the UV and IR branes.

First let us take a look at $J_{\gamma \gamma Z^{(\ell)}}$ in (\ref{anomalyJ1}) and $J_{ggZ^{(\ell)}}$ in (\ref{anomalyJ2}).
We decompose as
\begin{align}
\begin{pmatrix} \Tr \hat g^{Zuu}_{L \, \ell} - \Tr \hat g^{Zuu}_{R \, \ell}  \cr \mynoalign
\sum_{\tilde d = d, D_d} \Big( \Tr \hat g^{Z\tilde d \tilde d}_{L \, \ell} - \Tr \hat g^{Z \tilde d \tilde d}_{R \, \ell}  \Big) \cr \mynoalign
\Tr \hat g^{Zee}_{L \, \ell} - \Tr \hat g^{Zee}_{R \, \ell}  \end{pmatrix}
&= \begin{pmatrix} T^3_u F_{Z^{(\ell) }}^{u \, 1} \cr \mynoalign
T^3_d F_{Z^{(\ell) }}^{d\, 1} \cr \mynoalign
T^3_e F_{Z^{(\ell) }}^{e \, 1} \end{pmatrix}
- \sin^2 \theta_W^0  \begin{pmatrix} Q_u F_{Z^{(\ell) }}^{u \, 2}  \cr  \mynoalign
Q_d F_{Z^{(\ell) }}^{d \, 2}  \cr  \mynoalign
Q_e F_{Z^{(\ell) }}^{e \, 2}  \end{pmatrix} .
\label{Zelldecomposition}
\end{align}
Let us introduce
\begin{align}
&\begin{pmatrix} G_W[ (h^L, h^R,  {\hat h})_{\alpha};  ( f, g)_1, ( f, g)_2 ] \cr \mysnoalign
G_\gamma^u [ h_\gamma ;  ( f, g)_1, ( f, g)_2 ]  \cr \mysnoalign
G_\gamma^d [ h_\gamma ;  ( f, g, h, k)_1, ( f, g, h, k)_2 ] \end{pmatrix} \cr
\noalign{\kern 5pt}
&\quad
= \frac{k}{2} \sqrt{kL} \int_{-a}^{2L - a} dy \, e^{\sigma(y)} ~
\begin{pmatrix} {\cal K}_W[(h^L, h^R,  {\hat h})_{\alpha};  ( f, g)_1, ( f, g)_2 ](y) \cr \mysnoalign
{\cal K}_\gamma^u [ h_\gamma ;  ( f, g)_1, ( f, g)_2 ] (y) \cr
{\cal K}_\gamma^d [ h_\gamma ;  ( f, g, h, k)_1, ( f, g, h, k)_2 ] (y) \end{pmatrix}  ,  \cr \mysnoalign
\noalign{\kern 5pt}
&{\cal K}_W[(h^L, h^R,  {\hat h})_{\alpha};  ( f, g)_1, ( f, g)_2 ]  = h^{L*}_\alpha   f_1^*   f_2 +  h^{R*}_\alpha  g_1^*   g_2
+  \frac{i}{\sqrt{2}} \,  \hat h^*_\alpha  ( f_1^*  g_2 -  g_1^*  f_2 ) ,  \cr
\noalign{\kern 5pt}
&{\cal K}_\gamma^u [ h_\gamma ;  ( f, g)_1, ( f, g)_2 ] =  h_\gamma   (f_1^*   f_2 +   g_1^*   g_2) , \cr
\noalign{\kern 5pt}
&{\cal K}_\gamma^d [ h_\gamma ;  ( f, g, h, k)_1, ( f, g, h, k)_2 ] = 
 h_\gamma   (f_1^*   f_2 +   g_1^*   g_2  + h_1^* h_2 + k_1^* k_2 ) , 
\label{GWfunction2}
\end{align}
in the $y$ coordinate in the original gauge.  
Note that the functions ${\cal K}_W$, ${\cal K}_\gamma^u$, and ${\cal K}_\gamma^d$  in (\ref{GWfunction2}) 
are even under parity $y \go -y$ and periodic with a period $2 L$.  $a$ is arbitrary.
With Eq.\ (\ref{Zud1}) one finds that %, for instance, that
\begin{align}
&\begin{pmatrix} F_{Z^{(\ell) }}^{u \, 1}  \cr F_{Z^{(\ell) }}^{u \, 2}  \end{pmatrix}
= \frac{k}{2} \sqrt{kL} \cos \theta_W^0  \int_{-a}^{2L - a} dy \, e^{\sigma(y)} \sum_n \cr
\noalign{\kern 5pt}
&\quad \times
 \bigg\{ 
\begin{pmatrix}  {\cal K}_W[ (h^L, h^R, \hat h)^{su2}_{Z^{(\ell)}};  (f, g)^{u^{(n)}}_{L} , (f, g)^{u^{(n)}}_{L} ]  \cr
{\cal K}_\gamma^u [ h^{em}_{Z^{(\ell)}} ;   (f, g)^{u^{(n)}}_{L},   (f, g)^{u^{(n)}}_{L} ] \end{pmatrix} 
- \Big[ (f, g)_L \go (f, g)_R \Big] \bigg\} .
\label{FZell1}
\end{align}

For $J_{\gamma ZZ}$ in Eq.\ (\ref{anomalyJ4}) we decompose the coupling factors as 
\begin{align}
&\begin{pmatrix} \Tr \hat g^{Zuu}_{L \, 0}  \hat g^{Zuu}_{L \, 0}  - \Tr \hat g^{Zuu}_{R \, 0}  \hat g^{Zuu}_{R \, 0}  \cr
\mynoalign
\sum_{\tilde d_1 , \tilde d_2 = d, D_d} 
( \Tr \hat g^{Z \tilde d_1 \tilde d_2}_{L \, 0}   \hat g^{Z \tilde d_2 \tilde d_1}_{L \, 0} 
- \Tr \hat g^{Z \tilde d_1 \tilde d_2}_{R \, 0}   \hat g^{Z \tilde d_2 \tilde d_1}_{R \, 0}  )  \cr
\mynoalign
\Tr \hat g^{Zee}_{L \, 0}   \hat g^{Zee}_{L \, 0} 
- \Tr \hat g^{Zee}_{R\, 0}   \hat g^{Zee}_{R \, 0}  \end{pmatrix} \cr
\noalign{\kern 5pt}
&= 
\begin{pmatrix} (T^3_u  )^2 \,  F_{\gamma ZZ} ^{u \, 1}  \cr \mynoalign
(T^3_d  )^2 \,  F_{\gamma ZZ} ^{d \, 1}  \cr \mynoalign
(T^3_e  )^2 \,  F_{\gamma ZZ} ^{e \, 1} \end{pmatrix}
 - 2 \sin^2 \theta_W^0 
 \begin{pmatrix} T^3_u Q_u  F_{\gamma ZZ} ^{u \, 2}  \cr \mynoalign
T^3_d Q_d\,  F_{\gamma ZZ} ^{d \, 2}  \cr \mynoalign
T^3_e Q_e \,  F_{\gamma ZZ} ^{e \, 2} \end{pmatrix}
+ \sin^4 \theta_W^0 
 \begin{pmatrix} (Q_u)^2 F_{\gamma ZZ} ^{u \, 3}  \cr \mynoalign
(Q_d)^2 F_{\gamma ZZ} ^{d \, 3}  \cr \mynoalign
(Q_e)^2 F_{\gamma ZZ} ^{e \, 3} \end{pmatrix} .
\label{gammaZZdecomposition}
\end{align}
With Eq.\ (\ref{Zud1}) one finds that
\begin{align}
&\begin{pmatrix} F_{\gamma ZZ} ^{u \, 1} \cr F_{\gamma ZZ} ^{u \, 2}  \cr F_{\gamma ZZ} ^{u \, 3}  \end{pmatrix}
= \Big( \frac{k}{2} \sqrt{kL} \cos \theta_W^0  \Big)^2 \int_{-a}^{2L -a} dy_1  
\int_{-a}^{2L -a} dy_2  \, e^{\sigma (y_1) + \sigma (y_2)} \sum_{n,m=0}^\infty   \cr
\noalign{\kern 5pt}
&\hskip 1.cm
\times  \begin{pmatrix}
{\cal K}_{W, \, Z^{(0)}}^{u^{(n)}_{L}, u^{(m)}_{L} } (y_1)  {\cal K}_{W, \, Z^{(0)}}^{u^{(m)}_{L}, u^{(n)}_{L} } (y_2)
- {\cal K}_{W, \, Z^{(0)}}^{u^{(n)}_{R}, u^{(m)}_{R} } (y_1)  {\cal K}_{W, \, Z^{(0)}}^{u^{(m)}_{R}, u^{(n)}_{R} } (y_2)
\cr \mynoalign
{\cal K}_{W, \, Z^{(0)}}^{u^{(n)}_{L}, u^{(m)}_{L} } (y_1)  {\cal K}_{\gamma, \, Z^{(0)}}^{u^{(m)}_{L}, u^{(n)}_{L} } (y_2)
- {\cal K}_{W, \, Z^{(0)}}^{u^{(n)}_{R}, u^{(m)}_{R} } (y_1)  {\cal K}_{\gamma, \, Z^{(0)}}^{u^{(m)}_{R}, u^{(n)}_{R} } (y_2)
\cr \mynoalign
{\cal K}_{\gamma, \, Z^{(0)}}^{u^{(n)}_{L}, u^{(m)}_{L} } (y_1)  {\cal K}_{\gamma, \, Z^{(0)}}^{u^{(m)}_{L}, u^{(n)}_{L} } (y_2)
- {\cal K}_{\gamma, \, Z^{(0)}}^{u^{(n)}_{R}, u^{(m)}_{R} } (y_1)  {\cal K}_{\gamma, \, Z^{(0)}}^{u^{(m)}_{R}, u^{(n)}_{R} } (y_2)
\end{pmatrix} , \cr
\noalign{\kern 5pt}
&{\cal K}_{W, \, Z^{(0)}}^{u^{(n)}_{L/R}, u^{(m)}_{L/R} } (y) = 
{\cal K}_W[ (h^L, h^R, \hat h)^{su2}_{Z^{(0)}};  (f, g)^{u^{(n)}}_{L/R} , (f, g)^{u^{(m)}}_{L/R} ] (y)  , \cr
\noalign{\kern 5pt}
&{\cal K}_{\gamma, \, Z^{(0)}}^{u^{(n)}_{L/R}, u^{(m)}_{L/R} } (y) = 
{\cal K}_\gamma^u[ h^{em}_{Z^{(0)}};  (f, g)^{u^{(n)}}_{L/R} , (f, g)^{u^{(m)}}_{L/R} ] (y) .
\end{align}

At this stage we recognize that there are two ways to evaluate $(F_{Z^{(\ell)}}^{u \, 1}, F_{Z^{(\ell)}}^{u \, 2})$
and $(F_{\gamma ZZ} ^{u \, 1}, F_{\gamma ZZ} ^{u \, 2}, F_{\gamma ZZ} ^{u \, 3})$.

\noindent
%\quad 
$\underline{\hbox{Method 1}}$  First integrate over $y$ or $(y_1, y_2)$.  
Then do the KK sum $\sum_{n=0}^\infty$ or $\sum_{n,m=0}^\infty$.

\noindent
%\quad 
$\underline{\hbox{Method 2}}$  First do the KK sum $\sum_{n=0}^\infty$ or $\sum_{n,m=0}^\infty$. 
Then integrate over $y$ or $(y_1, y_2)$.

\noindent
Method 1 corresponds to evaluating each gauge coupling first and do the infinite KK sum. 
It was shown in Ref.\ \cite{AnomalyFlow2} that Method 2 leads to astonishing formulas in an $SU(2)$ model.

In performing the KK summation $\sum_{n=0}^\infty$ we encounter
\begin{align}
&\begin{pmatrix} A_{L/R}^{uu}  \cr B_{L/R}^{uu} \cr C_{L/R}^{uu} \cr D_{L/R}^{uu} \end{pmatrix}  (y_j, y_k)
= \sum_{n=0}^\infty \begin{pmatrix} f_{L/R}^{u^{(n)}} (y_j) f_{L/R}^{u^{(n)}} (y_k)^* \cr
g_{L/R}^{u^{(n)}} (y_j) g_{L/R}^{u^{(n)}} (y_k)^* \cr f_{L/R}^{u^{(n)}} (y_j) g_{L/R}^{u^{(n)}} (y_k)^* \cr 
g_{L/R}^{u^{(n)}} (y_j) f_{L/R}^{u^{(n)}} (y_k)^* \end{pmatrix} .
\label{completeness1}
\end{align}
Wave functions $f_{L}^{u^{(n)}}$ and $g_{R}^{u^{(n)}}$ are parity even, whereas  $f_{R}^{u^{(n)}}$ and 
$g_{L}^{u^{(n)}}$ are parity odd.
The KK sum  in (\ref{completeness1}) gives completeness relations on orbifolds.  
We shall confirm below by numerical evaluation that
\begin{align}
&F_{Z^{(\ell)}}^{\alpha \, j} = F_{Z^{(\ell)}}^{ j}  ~,  F_{\gamma ZZ} ^{\alpha \, k} = F_{\gamma ZZ}^{k} ~,  \cr
\noalign{\kern 5pt}
& (\alpha = u,d, c, s, t , b ~{\rm and}~e, \mu, \tau) ~.
\label{Fidentity1}
\end{align}
In other words the quantities $F_{Z^{(\ell)}}^{\alpha \, j} , F_{\gamma ZZ} ^{\alpha \, k}$ do not depend on 
the bulk mass parameter of fermions.
With this fact in mind one can evaluate the $(A_{L/R} ,B_{L/R}, C_{L/R}, D_{L/R})$ functions in Eq.\ (\ref{completeness1})
in the case of a vanishing bulk mass parameter $c=0$ for which all wave functions are expressed in terms of
trigonometric functions.  The derivation was given in Ref.\ \cite{AnomalyFlow2} and also is explained in Appendix B.
One finds
\begin{align}
&\underline{{\rm for~} c=0} \cr
&A_L(y,y') = B_R(y,y') = \frac{e^{-\sigma (y)}}{k} \big\{ \delta_{2L} (y - y') + \delta_{2L} (y + y') \big\} , \cr
\noalign{\kern 5pt}
&A_R(y,y') = B_L (y,y') = \frac{e^{-\sigma (y)}}{k} \big\{ \delta_{2L} (y - y') - \delta_{2L} (y + y') \big\} , \cr
\noalign{\kern 5pt}
&C_{L/R} (y,y') = D_{L/R} (y,y') = 0 ~, \cr
\noalign{\kern 5pt}
&\delta_L (x) = \sum_{n=-\infty}^\infty \delta (x - nL) ~.
\label{ABsumrelation1}
\end{align}

With the aid of the relations (\ref{Fidentity1}) and (\ref{ABsumrelation1}) one finds for $F_{Z^{(\ell)}}^{j}$ that
\begin{align}
F_{Z^{(\ell)}}^{ 1}  &= \frac{k}{2} \sqrt{kL} \cos \theta_W^0  \int_{-a}^{2L -a} dy \, e^{\sigma (y) } \cr
\noalign{\kern 5pt}
&\quad \times  \Big\{ h^{L, su2}_{Z^{(\ell)}}  (y) (A_L - A_R)(y,y) + h^{R, su2}_{Z^{(\ell)}}  (y) (B_L - B_R)(y,y) \Big\}  \cr
\noalign{\kern 5pt}
&= \sqrt{kL} \cos \theta_W^0  \int_{-a}^{2L -a} dy \, 
\Big\{  h^{L, su2}_{Z^{(\ell)}}  (y) -  h^{R, su2}_{Z^{(\ell)}}  (y) \Big\} \delta_{2L} (2y) \cr
\noalign{\kern 5pt}
&= \frac{1}{2}  \sqrt{kL} \cos \theta_W^0  \Big[  h^{L, su2}_{Z^{(\ell)}}  (0) -  h^{R, su2}_{Z^{(\ell)}}  (0) 
+  h^{L, su2}_{Z^{(\ell)}}  (L) -  h^{R, su2}_{Z^{(\ell)}}  (L) \Big]  , \cr
\noalign{\kern 5pt}
F_{Z^{(\ell)}}^{2}  &= \frac{k}{2} \sqrt{kL} \cos \theta_W^0  \int_{-a}^{2L -a} dy \, e^{\sigma (y) } 
h^{em}_{Z^{(\ell)}} (y) (A_L + B_L - A_R - B_R) (y,y) \cr
\noalign{\kern 5pt}
&= 0 ~.
\label{Fidentity2}
\end{align}
Similarly for $F_{\gamma ZZ}^{k} $ we find that
\begin{align}
F_{\gamma ZZ} ^{1} &= \Big( \frac{k}{2} \sqrt{kL} \cos \theta_W^0  \Big)^2 \int_{-a}^{2L -a} dy_1  
\int_{-a}^{2L -a} dy_2  \, e^{\sigma (y_1) + \sigma (y_2)} \cr
\noalign{\kern 5pt}
&%\hskip 1.cm
\times
\Big(  h^{L, su2}_1 h^{L, su2}_2 \big\{ A_L^{2,1}  A_L^{1,2}   - A_R^{2,1}  A_R^{1,2} \big\}  
+  h^{R, su2}_1 h^{R, su2}_2 \big\{ B_L^{2,1}  B_L^{1,2}   - B_R^{2,1}  B_R^{1,2} \big\}   \cr
\noalign{\kern 5pt}
&\hskip 1.cm 
+ \frac{1}{2} \hat h^{su2}_1 \hat h^{su2}_2
\big\{ A_L^{2,1}  B_L^{1,2}  + B_L^{2,1}  A_L^{1,2}   - A_R^{2,1}  B_R^{1,2} - B_R^{2,1}  A_R^{1,2}  \big\} \Big)  \cr
\noalign{\kern 5pt}
&= \Big( \frac{k}{2} \sqrt{kL} \cos \theta_W^0  \Big)^2 \int_{-a}^{2L -a} dy_1  
\int_{-a}^{2L -a} dy_2  \, e^{\sigma (y_1) + \sigma (y_2)} \cr
\noalign{\kern 5pt}
&\hskip 1.cm
\times
\big(  h^{L, su2}_1 h^{L, su2}_2 - h^{R, su2}_1 h^{R, su2}_2 \big) 
\big\{ A_L^{2,1}  A_L^{1,2}   - A_R^{2,1}  A_R^{1,2} \big\}  \cr
\noalign{\kern 5pt}
&= kL \cos^2  \theta_W^0   \int_{-a}^{2L -a} dy_1  dy_2 \, % \int_{-a}^{2L -a} dy_2 
\big(  h^{L, su2}_1 h^{L, su2}_2 - h^{R, su2}_1 h^{R, su2}_2 \big)  \delta_{2L} (y_1 - y_2) \delta_{2L} (y_1 +  y_2) \cr
\noalign{\kern 5pt}
&= \frac{kL}{2} \cos^2  \theta_W^0  \Big[ h^{L, su2}_{Z^{(0)}}  (0)^2 - h^{R, su2}_{Z^{(0)}}  (0)^2 
+ h^{L, su2}_{Z^{(0)}}  (L)^2 - h^{R, su2}_{Z^{(0)}}  (L)^2 \Big] ~, \cr
\noalign{\kern 5pt}
F_{\gamma ZZ} ^{2} &=  \frac{kL}{2} \cos^2  \theta_W^0 
\Big[ \big\{ h^{L, su2}_{Z^{(0)}} - h^{R, su2}_{Z^{(0)}} \big\}  h^{em}_{Z^{(0)}} \big|_{y=0}
+ \big\{ h^{L, su2}_{Z^{(0)}}- h^{R, su2}_{Z^{(0)}} \big\}  h^{em}_{Z^{(0)}} \big|_{y=L} \Big]  \cr
\noalign{\kern 5pt}
&= F_{\gamma ZZ} ^{1} ~, \cr
\noalign{\kern 5pt}
F_{\gamma ZZ} ^{3} &= 0 ~,
\label{Fidentity3}
\end{align}
where $h^{L, su2}_j =  h^{L, su2}_{Z^{(0)}}  (y_j)$,  $A_{L/R}^{j, k} = A_{L/R} (y_j, y_k)$ etc.
The equality $F_{\gamma ZZ} ^{2}  = F_{\gamma ZZ} ^{1}$ follows as 
$h^{em}_{Z^{(0)}} =  h^{L, su2}_{Z^{(0)}} + h^{R, su2}_{Z^{(0)}}$. 
We stress that $F$'s in Eqs.\ (\ref{Fidentity2}) and (\ref{Fidentity3}) are expressed in terms of
the values of the $Z$ wave functions at $y=0$ and $y=L$.

As mentioned above, the independence of $F$'s on the fermion species, namely the relations (\ref{Fidentity1}),
is numerically checked.  Let us define $F_{Z^{(\ell)}}^{\alpha \, j } |_{n_{\rm max}}$ and  
 $F_{\gamma ZZ } ^{\alpha \, k} |_{n_{\rm max}}$ by
\begin{align}
&\sum_{n=0}^{n_\max} \Big\{ (\hat g^{Zuu}_{L \, \ell})_{nn} - (\hat g^{Zuu}_{R \, \ell})_{nn}  \Big\} 
= T^3_u F_{Z^{(\ell) }}^{u \, 1} |_{n_\max} - \sin^2 \theta_W^0 Q_u  F_{Z^{(\ell) }}^{u \, 2} |_{n_\max} ~, \cr
\noalign{\kern 5pt}
&\sum_{n, m=0}^{n_\max} \Big\{ (\hat g^{Zuu}_{L \, 0})_{nm} (\hat g^{Zuu}_{L \, 0})_{mn}  
- (\hat g^{Zuu}_{R \, 0})_{nm} (\hat g^{Zuu}_{R \, 0})_{mn}   \Big\}   \cr
\noalign{\kern 5pt}
&= (T^3_u  )^2 \,  F_{\gamma ZZ} ^{u \, 1}  |_{n_\max}
 - 2 \sin^2 \theta_W^0  T^3_u Q_u  F_{\gamma ZZ} ^{u \, 2}  |_{n_\max} 
 + \sin^4 \theta_W^0   (Q_u)^2 F_{\gamma ZZ} ^{u \, 3}  |_{n_\max}  ~.
 \label{Fnmax1}
\end{align}
In other words $F |_{n_\max}$ is defined by doing the KK sum up to the KK level $n_\max$.
The relations  (\ref{Fidentity1}) are confirmed by showing
\begin{align}
&\lim_{n_\max \go \infty} F_{Z^{(\ell)}}^{\alpha \, j}  |_{n_\max} = F_{Z^{(\ell)}}^{ j}  ~,  \cr
\noalign{\kern 5pt}
&\lim_{n_\max \go \infty} F_{\gamma ZZ} ^{\alpha \, k}  |_{n_\max} = F_{\gamma ZZ}^{k} ~.
\label{FidentityX1}
\end{align}
The result for $F_{Z^{(0)}}^{\alpha \, 1}$ ($\alpha = u, d, e, t, b, \tau$) is tabulated in Table  \ref{Table:ggZ0}.
It is seen that the convergence of $F_{Z^{(0)}}^{\alpha \, 1} |_{n_\max}$ to  $F_{Z^{(0)}}^{1} $ is fast.  
The result for $F_{Z^{(1)}}^{\alpha \, 1}$  is given in Table  \ref{Table:ggZ1} in Appendix C. 
There is large parity violation in the $Z^{(1)}$ couplings of the low-lying KK modes %($n=0,1,2$ modes)  
of fermions.  The convergence of $F_{Z^{(1)}}^{\alpha \, 1} |_{n_\max}$ to  $F_{Z^{(1)}}^{1} $ is
slower than that of $F_{Z^{(0)}}^{\alpha \, 1} |_{n_\max}$ to  $F_{Z^{(0)}}^{1} $.

\begin{table}[tbh]
\renewcommand{\arraystretch}{1.2}
\begin{center}
\caption{$F_{Z^{(0)}}^{\alpha \, 1}  |_{n_\max} $ ($\alpha = u, d, e, t, b, \tau$)  is tabulated
for $\theta_H=0.1$ and $m_\KK = 13\,$TeV.
The bottom row $n_\max = \infty$ represents the value of the formula in Eq.\ (\ref{Fidentity2}).
}
\vskip 10pt
\begin{tabular}{ccccccc}
\hline 
$n_\max$ & $F_{Z^{(0)}}^{u \, 1}$ & $F_{Z^{(0)}}^{d \, 1}$ &$F_{Z^{(0)}}^{e \, 1}$  
   & $F_{Z^{(0)}}^{t \, 1}$ & $F_{Z^{(0)}}^{b\, 1}$ &$F_{Z^{(0)}}^{\tau \, 1}$  \\
\noalign{\kern 2pt}
\hline
0  &$0.997688$ &$0.997688$ &$0.997691$  &$0.997671$ &$0.997671$ &$0.997684$ \\
6  &$0.997659$ &$0.997659$ &$0.997661$ &$0.997653$ &$0.997653$ &$0.997657$ \\
12 &$0.997655$ &$0.997655$ &$0.997656$ &$0.997651$ &$0.997651$ &$0.997653$\\
18 &$0.997653$ &$0.997653$ &$0.997654$ &$0.997651$ &$0.997651$ &$0.997652$\ \\
24&$0.997652$  &$0.997652$ &$0.997653$ &$0.997650$ &$0.997650$ &$0.997651$\\
\noalign{\kern 5pt}
\hline
$\infty$ &  & &$0.997649$ & && \\
\hline 
\end{tabular}
\label{Table:ggZ0}
\end{center}
\end{table}

%%%%%%%%%%%%%%%%%%%%

The result for $F_{\gamma ZZ}^{\alpha \, 1}$  ($\alpha = u, d, e, t, b, \tau$)  is tabulated in Table \ref{Table:gammaZZ1}.
It is seen that $F_{\gamma ZZ}^{\alpha \, 1} |_{n_\max}$ converges to $F_{\gamma ZZ}^{1} $ as
$n_\max \go \infty$.
Similarly one can check that $F_{\gamma ZZ}^{\alpha \, 2} |_{n_\max \go \infty} = F_{\gamma ZZ}^{2} = F_{\gamma ZZ}^{1}$.
See Table  \ref{Table:gammaZZ2} in Appendix C.
We observe  that
\begin{align}
&F_{\gamma ZZ}^{\alpha \, 1} |_{n_\max} <  F_{\gamma ZZ}^{1} = F_{\gamma ZZ}^{2} 
< F_{\gamma ZZ}^{\alpha \, 2} |_{n_\max} 
\quad \hbox{for } n_\max < \infty .
\label{FgammaZZrelation}
\end{align}

\begin{table}[tbh]
\renewcommand{\arraystretch}{1.2}
\begin{center}
\caption{$F_{\gamma ZZ}^{\alpha \, 1}  |_{n_\max} $ ($\alpha = u, d, e, t, b, \tau$)  is tabulated
for $\theta_H=0.1$ and $m_\KK = 13\,$TeV.
The bottom row $n_\max = \infty$ represents the value of the formula in Eq.\ (\ref{Fidentity3}).
}
\vskip 10pt
\begin{tabular}{ccccccc}
%\begin{tabular}{|c|c|c|c|c|c|}
\hline 
$n_\max$ & $F_{\gamma ZZ}^{u \, 1}$ & $F_{\gamma ZZ}^{d \, 1}$ &$F_{\gamma ZZ}^{e \, 1}$  
   & $F_{\gamma ZZ}^{t \, 1}$ & $F_{\gamma ZZ}^{b\, 1}$ &$F_{\gamma ZZ}^{\tau \, 1}$  \\
\noalign{\kern 2pt}
\hline
0  &$0.995381$ &$0.995381$ &$0.995386$  &$0.996691$ &$0.995348$ &$0.995374$ \\
6 &$0.997479$ &$0.997479$ &$0.997433$ &$0.997692$ &$0.997679$ &$0.997539$ \\
12&$0.997628$ &$0.997628$ &$0.997601$ &$0.997742$ &$0.997734$ &$0.997661$\\
18 &$0.997682$ &$0.997682$ &$0.997663$ &$0.997759$ &$0.997753$ &$0.997704$\ \\
24 &$0.997709$  &$0.997709$ &$0.997695$ &$0.997768$ &$0.997762$ &$0.997726$\\
\noalign{\kern 5pt}
\hline
$\infty$ &  & &$0.997796$ & && \\
\hline 
\end{tabular}
\label{Table:gammaZZ1}
\end{center}
\end{table}

For  $J_{\gamma WW}$ in (\ref{anomalyJ3}) relevant quantities are 
\begin{align}
F_{\gamma WW}^{ud} &= \sum_{\tilde d = d, D_d}
( \Tr \hat g^{Wu \tilde d}_{L \, 0}  \hat g^{W^\dagger \tilde d u}_{L \, 0}  
- \Tr \hat g^{Wu \tilde d}_{R \, 0}  \hat g^{W^\dagger \tilde d u}_{R \, 0}  )  ~, \cr
\noalign{\kern 5pt}
F_{\gamma WW}^{\nu_e e} &=  \sum_{a=+,-}  
(\Tr \hat g^{W^\dagger e \nu_e^a}_{L \, 0}   \hat g^{W\nu_e^a e}_{L \, 0} 
- \Tr \hat g^{W^\dagger e \nu_e^a}_{R\, 0}   \hat g^{W\nu_e^a e}_{R \, 0}  ) ~.
\label{FWW}
\end{align}
Again one can check numerically that
\begin{align}
&F_{\gamma WW}^{\alpha } = F_{\gamma WW}  ~, \cr
\noalign{\kern 5pt}
&\alpha = ud, ~ cs, ~ tb,  ~ \nu_e e,  ~ \nu_\mu \mu, ~ \nu_\tau \tau ~. 
\label{Fidentity4}
\end{align}
Analytic form of $F_{\gamma WW} $ is found for the limiting case when the bulk mass parameter $c=0$ and 
brane interactions are absent.  In this case $u$ and $d$ quarks are degenerate and 
$g^{Wu D_d}_{L/R \, 0}  =0$.
Using $\hat g^{W ud}_{L/R\, 0 n m}  = G_W[ (h^L, h^R, {\hat h})_{W^{(0)}}; (f, g)^{u^{(n)}}_{L/R} , (f, g)^{d^{(m)}}_{L/R} ] $
and doing $\Tr$ (the KK sum $\sum_{n,m}$) first, one finds, as in Eq.\ (\ref{Fidentity3}),   that
\begin{align}
F_{\gamma WW}  &=  \Big( \frac{k}{2} \sqrt{kL} \Big)^2 \int_{-a}^{2L -a} dy_1  
\int_{-a}^{2L -a} dy_2  \, e^{\sigma (y_1) + \sigma (y_2)} \cr
\noalign{\kern 5pt}
&%\hskip 1.cm
\times
\Big(  h^{L \, *}_1 h^{L}_2 \big\{ A_L^{2,1}  A_L^{1,2}   - A_R^{2,1}  A_R^{1,2} \big\}  
+  h^{R \, *}_1 h^{R}_2 \big\{ B_L^{2,1}  B_L^{1,2}   - B_R^{2,1}  B_R^{1,2} \big\}   \cr
\noalign{\kern 5pt}
&\hskip 1.cm 
+ \frac{1}{2} \hat h_1^*  \hat h_2
\big\{ A_L^{2,1}  B_L^{1,2}  + B_L^{2,1}  A_L^{1,2}   - A_R^{2,1}  B_R^{1,2} - B_R^{2,1}  A_R^{1,2}  \big\} \Big)  \cr
\noalign{\kern 5pt}
&= \frac{kL}{2} \Big[ \big| h^{L}_{W^{(0)}}  (0) \big|^2 - \big| h^{R}_{W^{(0)}} (0) \big|^2 
+ \big| h^{L}_{W^{(0)}} (L) \big|^2 - \big| h^{R}_{W^{(0)}} (L) \big|^2 \Big] ~,
\label{Fidentity5}
\end{align}
where $h^{L}_j =  h^{L}_{W^{(0)}}  (y_j)$ etc.
As in Eq.\ (\ref{Fnmax1}) we define $F_{\gamma WW}^\alpha |_{n_\max}$ ($\alpha = ud, \nu_e e, ~$etc)
by doing the KK sum up to the KK level $n_\max$.
The relation (\ref{Fidentity4}) is confirmed by showing
\begin{align}
&\lim_{n_\max \go \infty} F_{\gamma WW}^\alpha |_{n_\max} = F_{\gamma WW} ~.
\label{FidentityX2}
\end{align}
The result for $F_{\gamma WW}^\alpha$ ($\alpha = ud, \nu_e e, tb, \nu_\tau \tau$) is tabulated in Table \ref{Table:gammaWW}
in Appendix C.

Arguments are generalized to $J_{ZZZ}$ in (\ref{anomalyJ6}).  For the $u$ loop we decompose as 
\begin{align}
&\Tr \hat g^{Zuu}_{L \, 0}  \hat g^{Zuu}_{L \, 0}  \hat g^{Zuu}_{L \, 0} 
- \Tr \hat g^{Zuu}_{R \, 0}  \hat g^{Zuu}_{R \, 0}   \hat g^{Zuu}_{R \, 0}  \cr
\noalign{\kern 5pt}
&= (T^3_u )^3 F_{ZZZ}^{u \,1} - 3 \sin^2 \theta_W^0 (T^3_u )^2 Q_u  F_{ZZZ}^{u \,2} \cr
\noalign{\kern 5pt}
&\hskip 1.cm 
+ 3 \sin^4 \theta_W^0 T^3_u  (Q_u )^2 F_{ZZZ}^{u \,3}  - \sin^6 \theta_W^0  (Q_u )^3 F_{ZZZ}^{u \,4} ~.
\label{ZZZdecomposition1}
\end{align}
With similar decompositions one can check numerically that
\begin{align}
F_{ZZZ}^{u \,j}  = F_{ZZZ}^{d \, j}  =  F_{ZZZ}^{e \, j }  \equiv F_{ZZZ}^j  ~, ~~~ F_{ZZZ}^{\nu_e \, 1}  = F_{ZZZ}^1 ~.
\label{ZZZidentity}
\end{align}
Analytic form of $F_{ZZZ}^j $ is conveniently found for $c=0$.
One finds for $F_{ZZZ}^1 $ that 
\begin{align}
F_{ZZZ}^1 &= 
\Big( \frac{k}{2} \sqrt{kL} \cos \theta_W^0  \Big)^3 \int_{-a}^{2L -a} dy_1   \int_{-a}^{2L -a} dy_2  
\int_{-a}^{2L -a} dy_3  \, e^{\sigma (y_1) + \sigma (y_2) + \sigma (y_3)} \cr
\noalign{\kern 5pt}
&%\hskip 1.cm
\times
\Big( h^{L, su2}_1 h^{L, su2}_2 h^{L, su2}_3  \big\{ A_L^{1,2}  A_L^{2,3}  A_L^{3,1}  - A_R^{1,2}  A_R^{2,3} A_R^{3,1}  \big\}  \cr
\noalign{\kern 5pt}
&\quad 
+ h^{R, su2}_1 h^{R, su2}_2 h^{R, su2}_3  \big\{ B_L^{1,2}  B_L^{2,3}  B_L^{3,1}  - B_R^{1,2}  B_R^{2,3} B_R^{3,1}\big\}   \cr
\noalign{\kern 5pt}
&\quad
+ \frac{3}{2}  h^{L, su2}_1  \hat h^{su2}_2 \hat h^{su2}_3
\big\{ A_L^{1,2} B_L^{2,3}  A_L^{3,1}  - A_R^{1,2}  B_R^{2,3} A_R^{3,1}  \big\} \cr
\noalign{\kern 5pt}
&\quad
+ \frac{3}{2}  h^{R, su2}_1  \hat h^{su2}_2 \hat h^{su2}_3
\big\{ B_L^{1,2}  A_L^{2,3}  B_L^{3,1}  - B_R^{1,2}  A_R^{2,3} B_R^{3,1}\big\}  \Big)  ~.
\label{FZZZ1}
\end{align}
Inserting the expressions in Eq.\ (\ref{ABsumrelation1}) into the above, one finds that
\begin{align}
&F_{ZZZ}^1 = \frac{1}{4}
\Big( \sqrt{kL} \cos \theta_W^0  \Big)^3 \int_{-a}^{2L -a} dy_1   \int_{-a}^{2L -a} dy_2  \int_{-a}^{2L -a} dy_3  \, \cr
\noalign{\kern 5pt}
&\qquad
\times \Big( h^{L, su2}_1 h^{L, su2}_2 h^{L, su2}_3 - h^{R, su2}_1 h^{R, su2}_2 h^{R, su2}_3 \Big) \cr
\noalign{\kern 5pt}
&\qquad
\times \big[ \delta_{2L} (y_1 - y_2) \delta_{2L} (y_2 - y_3) \delta_{2L} (y_3 + y_1) \cr
\noalign{\kern 5pt}
&\qquad
+ \delta_{2L} (y_1 - y_2) \delta_{2L} (y_2 + y_3) \delta_{2L} (y_3 - y_1) \cr
\noalign{\kern 5pt}
&\qquad
+ \delta_{2L} (y_1 +y_2) \delta_{2L} (y_2 - y_3) \delta_{2L} (y_3 - y_1) \cr
\noalign{\kern 5pt}
&\qquad
+ \delta_{2L} (y_1 +y_2) \delta_{2L} (y_2 + y_3) \delta_{2L} (y_3 + y_1)  \big] \cr
\noalign{\kern 5pt}
&=  \frac{1}{2} \Big( \sqrt{kL} \cos \theta_W^0  \Big)^3 
\Big[ h^{L, su2}_{Z^{(0)}}  (0)^3 - h^{R, su2}_{Z^{(0)}}  (0)^3 
+ h^{L, su2}_{Z^{(0)}}  (L)^3 - h^{R, su2}_{Z^{(0)}}  (L)^3 \Big] ~.
\label{Fidentity6}
\end{align}
Similarly one finds that
\begin{align}
F_{ZZZ}^2 &=  \frac{1}{2} \Big( \sqrt{kL} \cos \theta_W^0  \Big)^3 
\Big[ \big\{ h^{L, su2}_{Z^{(0)}}  (0)^2 - h^{R, su2}_{Z^{(0)}}  (0)^2 \big\} h^{em}_{Z^{(0)}}  (0) \cr
\noalign{\kern 5pt}
&\hskip 3.cm
+ \big\{ h^{L, su2}_{Z^{(0)}}  (L)^2 - h^{R, su2}_{Z^{(0)}}  (L)^2 \big\}   h^{em}_{Z^{(0)}}  (L) \Big] ~,  \cr
\noalign{\kern 5pt}
F_{ZZZ}^3 & = \frac{1}{2} \Big( \sqrt{kL} \cos \theta_W^0  \Big)^3 
\Big[ \big\{ h^{L, su2}_{Z^{(0)}}  (0) - h^{R, su2}_{Z^{(0)}} (0) \big\} h^{em}_{Z^{(0)}}  (0)^2 \cr
\noalign{\kern 5pt}
&\hskip 3.cm
+ \big\{ h^{L, su2}_{Z^{(0)}} (L) - h^{R, su2}_{Z^{(0)}} (L) \big\}   h^{em}_{Z^{(0)}} (L)^2 \Big] \cr
\noalign{\kern 5pt}
&= F_{ZZZ}^2 ~, \cr
\noalign{\kern 5pt}
F_{ZZZ}^4 &=  0 ~.
\label{Fidentity7}
\end{align}
For $\theta_H = 0.1$ and $m_\KK = 13\,$TeV, $F_{ZZZ}^1 =0.996703$ and $F_{ZZZ}^2 = F_{ZZZ}^3 = 0.997943$.
The relation (\ref{ZZZidentity}) is checked numerically.  By defining  $F_{ZZZ}^{\alpha \, j} |_{n_\max}$ as before,
one confirms that $\lim_{n_\max \go \infty} F_{ZZZ}^{\alpha \, j} |_{n_\max} = F_{ZZZ}^{j} $.  
The result for $F_{ZZZ}^{\alpha \, 1}  |_{n_\max} $ ($\alpha = u, d, \nu_e, e, t, b, \nu_\tau, \tau$)  
is tabulated in Table \ref{Table:ZZZ1} in Appendix C.
The result for $F_{ZZZ}^{\alpha \, 2}  |_{n_\max} $ ($\alpha = u, d,  e, t, b, \tau$)  
is tabulated in Table \ref{Table:ZZZ2} in Appendix C.

Similarly  $\lim_{n_\max \go \infty} F_{ZZZ}^{\alpha \, 3} |_{n_\max} = F_{ZZZ}^{3}$ is checked.  
It is found that 
\begin{align}
&F_{ZZZ}^{\alpha \, 2} |_{n_\max} <  F_{ZZZ}^{2} = F_{ZZZ}^{3} 
< F_{ZZZ}^{\alpha \, 3} |_{n_\max} 
\quad \hbox{for } n_\max < \infty .
\label{FZZrelation}
\end{align}
The result for $F_{ZZZ}^{\alpha \, 4}  |_{n_\max} $ ($\alpha = u, d,  e, t, b, \tau$)  
is tabulated in Table \ref{Table:ZZZ4} in Appendix C.
We note that in evaluating $F_{ZZZ}^{\alpha \, 4}  |_{n_\max} $ ($\alpha = d,  s, b$)  
contributions coming from $D_d, D_s, D_b$ towers are very important.

For $J_{ZWW}$ in (\ref{anomalyJ5})  we decompose as 
\begin{align}
&\sum_{\tilde d = d, D_d}
( \Tr \hat g^{Wu \tilde d}_{L \, 0}  \hat g^{W^\dagger \tilde d u}_{L \, 0}  \hat g^{Zuu}_{L \, 0} 
- \Tr \hat g^{Wu \tilde d}_{R \, 0}  \hat g^{W^\dagger \tilde d u}_{R \, 0}   \hat g^{Zuu}_{R \, 0}  ) 
= T^3_u F_{ZWW}^{u \, 1} - \sin^2 \theta_W^0 Q_u F_{ZWW}^{u \, 2}  , \cr
\noalign{\kern 5pt}
& \sum_{\tilde d_1, \tilde d_2 = d, D_d}
( \Tr \hat g^{W^\dagger \tilde d_1 u}_{L \, 0}   \hat g^{Wu \tilde d_2}_{L \, 0}   \hat g^{Z \tilde d_2 \tilde d_1}_{L \, 0} 
- \Tr \hat g^{W^\dagger \tilde d_1 u}_{R \, 0}   \hat g^{Wu \tilde d_2}_{R \, 0}  \hat g^{Z \tilde d_2 \tilde d_1}_{R \, 0} ) 
= T^3_d F_{ZWW}^{d \, 1} - \sin^2 \theta_W^0 Q_d F_{ZWW}^{d \, 2}  , \cr
\noalign{\kern 5pt}
&\sum_{a, b=+,-}  (\Tr \hat g^{W  \nu_e^a e}_{L \, 0}   \hat g^{W^\dagger e \nu_e^b}_{L \, 0} \hat g^{Z  \nu_e^b  \nu_e^a}_{L \, 0} 
- \Tr  \hat g^{W\nu_e^a e}_{R \, 0}   \hat g^{W^\dagger e \nu_e^b}_{R\, 0} \hat g^{Z  \nu_e^b  \nu_e^a}_{R \, 0}  )  
= T^3_{\nu} F_{ZWW}^{\nu_e \, 1}  , \cr
\noalign{\kern 5pt}
&\sum_{a = +,-}  (\Tr \hat g^{W^\dagger e \nu_e^a}_{L \, 0}   \hat g^{W\nu_e^a e}_{L \, 0}   \hat g^{Zee}_{L \, 0}
- \Tr \hat g^{W^\dagger e \nu_e^a}_{R\, 0}   \hat g^{W\nu_e^a e}_{R \, 0}   \hat g^{Zee}_{R \, 0} ) 
=  T^3_e F_{ZWW}^{e \, 1} - \sin^2 \theta_W^0 Q_e F_{ZWW}^{e \, 2}  .
\label{ZWWdecomposition1}
\end{align}
We find numerically that
\begin{align}
&F_{ZWW}^{\alpha \, 1}  = F_{ZWW}^{1}  \quad {\rm for~} \alpha = u, d, \nu_e, e ~, \cr
&F_{ZWW}^{\beta\, 2}  = F_{ZWW}^{2}  \quad {\rm for~} \beta = u, d, e ~.
\label{ZWWidentity}
\end{align}
Analytic form of $F_{ZWW}^{j}$ is found, as in the case of $F_{\gamma WW}$, by considering
the case with vanishing bulk mass parameter and brane interactions.
Manipulations are similar to those for $F_{ZZZ}^{j}$.  One finds that
\begin{align}
F_{ZWW}^{1}  &=\frac{1}{2}  ( kL )^{3/2} \cos \theta_W^0  
\Big[ | h^{L}_{W^{(0)}} (0) |^2  h^{L, su2}_{Z^{(0)}} (0)  -  | h^{R}_{W^{(0)}} (0) |^2 h^{R, su2}_{Z^{(0)}} (0)\cr
\noalign{\kern 5pt}
&\hskip 2.8cm
+  | h^{L}_{W^{(0)}} (L) |^2  h^{L, su2}_{Z^{(0)}} (L)  -  | h^{R}_{W^{(0)}} (L) |^2 h^{R, su2}_{Z^{(0)}} (L) \Big] ,  \cr
\noalign{\kern 5pt}
F_{ZWW}^{2}  &=\frac{1}{2}  ( kL )^{3/2} \cos \theta_W^0  
\Big[  \Big( | h^{L}_{W^{(0)}} (0) |^2  - | h^{R}_{W^{(0)}} (0) |^2  \Big) h^{em}_{Z^{(0)}} (0) \cr
\noalign{\kern 5pt}
&\hskip 2.8cm
+  \Big( | h^{L}_{W^{(0)}} (L) |^2  - | h^{R}_{W^{(0)}} (L) |^2  \Big) h^{em}_{Z^{(0)}} (L)  \Big] .
\label{Fidentity8}
\end{align}
The relation (\ref{ZWWidentity}) is checked numerically.  By defining  $F_{ZWW}^{\alpha \, j} |_{n_\max}$ as before,
one confirms that $\lim_{n_\max \go \infty} F_{ZWW}^{\alpha \, j} |_{n_\max} = F_{ZWW}^{j} $.  
The results for $F_{ZWW}^{1}$  and $F_{ZWW}^{2}$ are tabulated in Table \ref{Table:ZWW1} and Table \ref{Table:ZWW2}
in Appendix C, respectively.

We have established holographic formulas (\ref{Fidentity2}), (\ref{Fidentity3}), (\ref{Fidentity5}), 
(\ref{Fidentity6}), (\ref{Fidentity7}) and (\ref{Fidentity8}) for chiral anomalies.
The $F$ factors are expressed in terms of the values of the $W$ and $Z$ wave functions at the UV and IR branes,
irrespective of fermion species running along triangular loops.  It is important to include the contributions of
all KK fermion modes.

\begin{figure}[tbh]
\centering
\includegraphics[height=45mm]{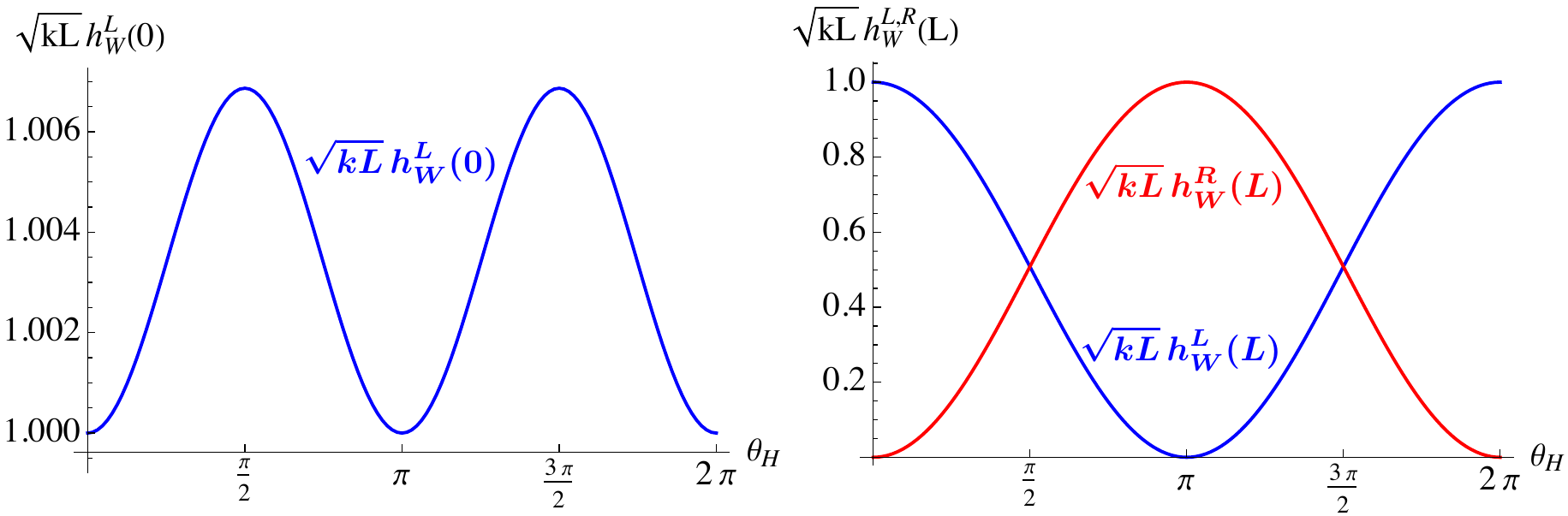}
\caption{The $\theta_H$-dependence of $\sqrt{kL} \, \{  h^L_W (0, \theta_H), h^L_W (L, \theta_H),  h^R_W (L, \theta_H) \}$ 
is shown.  $h^R_W (0, \theta_H) \sim 0$.
The behavior of $\cos \theta_W^0 \sqrt{kL} \,  h^{L/R, su2}_Z (y, \theta_H)$ at $y=0, L$ is similar to that of 
$\sqrt{kL} \,  h^{L/R}_W (y, \theta_H)$ at $y=0, L$.
}
\label{fig:hW}
\end{figure}

\begin{figure}[tbh]
\centering
\includegraphics[height=45mm]{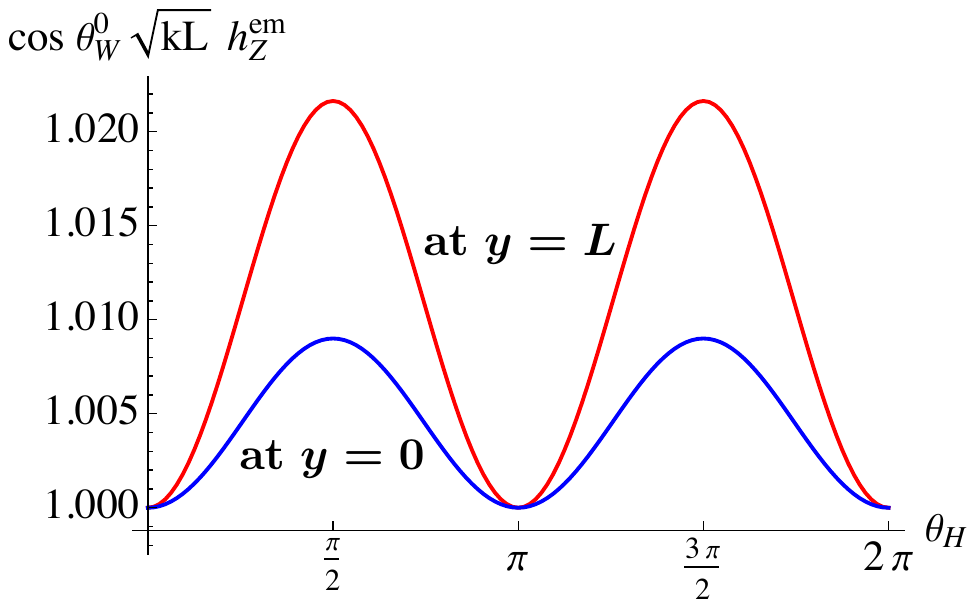}
\caption{The $\theta_H$-dependence of  $\cos \theta_W^0 \sqrt{kL} \,  \{ h^{em}_Z (0, \theta_H) , h^{em}_Z (L, \theta_H) \}$ is displayed.
}
\label{fig:hemZ}
\end{figure}

\begin{figure}[tbh]
\centering
\includegraphics[height=45mm]{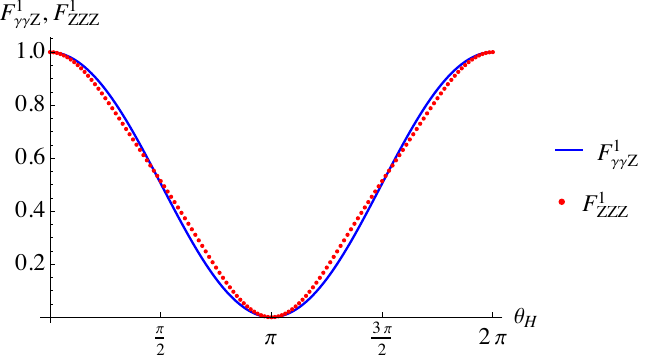}
\caption{The $\theta_H$-dependence of $F^1_{\gamma \gamma Z}  = F^1_{Z^{(0)}}$ and $F^1_{ZZZ} $ is displayed.
}
\label{fig:FZ}
\end{figure}

\begin{figure}[tbh]
\centering
\includegraphics[height=45mm]{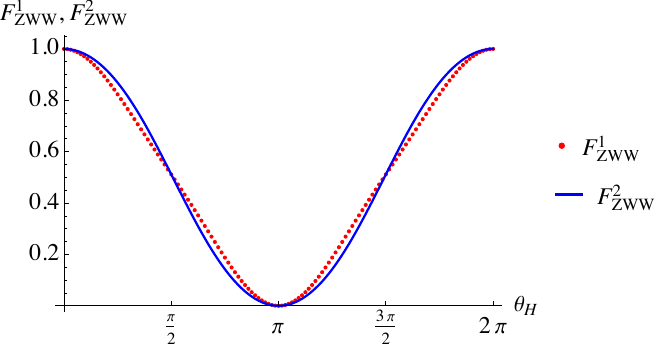}
\caption{The $\theta_H$-dependence of $F^1_{ZWW} $  and $F^2_{ZWW} $ is displayed.
}
\label{fig:FZWW}
\end{figure}

Anomalies flow with $\theta_H$.  The $\theta_H$-dependence of the anomalies can be easily seen 
from the holographic formulas derived above.  The wave functions of the $W$ and $Z$ bosons depend 
on $\theta_H$ and $z_L$  so that various $F$ factors are functions of $\theta_H$ and $z_L$; $F = F(\theta_H, z_L)$.
The $\theta_H$-dependence of the wave functions of $W$ and $Z$ at $y=0$ and $L$ with $z_L$ kept fixed 
is displayed in Figs.\ \ref{fig:hW} and \ref{fig:hemZ} .
Each anomaly coefficient is proportional to the $F$ factor.
The $\theta_H$-dependence of  $F^1_{\gamma \gamma Z}\equiv  F^1_{Z^{(0)}}$, $F^1_{ZZZ}$,
$F^1_{ZWW} $  and $F^2_{ZWW}$ is displayed in Fig.\ \ref{fig:FZ} and \ref{fig:FZWW}.
One sees that $F$'s smoothly change with $\theta_H$ in the RS space ($z_L > 1$).
These $F$'s  vanish at $\theta_H = \pi$ where quark and lepton multiplet fields become vector-like in gauge couplings.
In the flat spacetime limit, namely in the  $k \go 0$ and $z_L \go 1$ limit, the $F$ factors show the behavior of step functions
at $\theta_H = 0, \pm \onehalf \pi, \pm \pi$ where level crossing in the mass spectrum of gauge bosons
takes place, as was shown in the $SU(2)$ model in Ref.\ \cite{AnomalyFlow2}.

%%%%%%%%%%%%%%%%%%%%
%%%%%%%%%%%%%%%%%%%

%%%%%%%%%%%%%%%%%%%%
%%%%%%%%%%%%%%%%%%%

%%%%%%%%%%%%%%%%%%%%
%%%%%%%%%%%%%%%%%%%

%%%%%%%%%%%%%%%%%%%%
%%%%%%%%%%%%%%%%%%%

\section{Anomaly  cancellation} 

We have seen in the previous section that total chiral anomalies, which take into account the contributions of
all fermion KK modes, are expressed in terms of the values of wave functions of gauge bosons at the
UV and IR branes.  The magnitude of the total anomalies is independent of bulk mass parameters of
fermions running along internal triangular loops.
With this fact  the cancellation of gauge anomalies in the GUT-inspired GHU model is achieved as shown below.

Gauge anomalies for $\gamma \gamma Z^{(\ell)}$  and  $ggZ^{(\ell)}$  are proportional to 
 $J_{\gamma \gamma Z^{(\ell)}}$ and $J_{gg Z^{(\ell)}}$ in Eqs.\ (\ref{anomalyJ1}) and (\ref{anomalyJ2}), respectively.   
 With Eqs.\  (\ref{Fidentity1}) and (\ref{Fidentity2}) one finds that 
\begin{align}
J_{\gamma \gamma Z^{(\ell)}}  &=
\big\{ 3 [ (Q_u)^2 T^3_u + (Q_d)^2 T^3_d ] + (Q_e)^2 T^3_e \big\} \, F^1_{Z^{(\ell)}}  \cr
\noalign{\kern 5pt}
&=  0 ~, \cr
\noalign{\kern 5pt}
J_{gg Z^{(\ell)}}  &=
3  ( T^3_u + T^3_d )  \, F^1_{Z^{(\ell)}}  \cr
\noalign{\kern 5pt}
&= 0 ~.
\label{anom1}
\end{align}
For $\gamma Z Z$,  with Eqs.\  (\ref{Fidentity1}) and (\ref{Fidentity3}), one finds that
\begin{align}
J_{\gamma Z Z}  &=
\big\{ 3 [ Q_u  (T^3_u)^2 + Q_d (T^3_d)^2 ] + Q_e (T^3_e)^2  \big\} \, F^1_{\gamma ZZ}  \cr
\noalign{\kern 5pt}
&\quad - 2 \sin^2 \theta_W^0 \big\{ 3 [ (Q_u)^2 T^3_u +  (Q_d)^2 T^3_d  ] + (Q_e)^2 T^3_e  \big\} \, F^2_{\gamma ZZ} \cr
\noalign{\kern 5pt}
&=  0 ~.
\label{anom2}
\end{align}
For $\gamma WW$,  with Eqs.\  (\ref{Fidentity4}) and (\ref{Fidentity5}), 
\begin{align}
J_{\gamma WW}  &=
\big\{ 3 [ Q_u  + Q_d ] + Q_e  \big\} \, F_{\gamma WW}  \cr
\noalign{\kern 5pt}
&=  0 ~.
\label{anom3}
\end{align}
For $ZZZ$,  with Eqs.\  (\ref{ZZZidentity}), (\ref{Fidentity6})  and (\ref{Fidentity7}), 
\begin{align}
J_{ZZZ}  &=
\big\{ 3 [ (T^3_u)^3 +  (T^3_d)^3] +  (T^3_\nu)^3  + (T^3_e)^3  \big\} \, F^1_{ZZZ}  \cr
\noalign{\kern 5pt}
&\quad - 3 \sin^2 \theta_W^0 \big\{ 3 [ Q_u (T^3_u)^2 +  Q_d (T^3_d )^2 ] + Q_e (T^3_e)^2  \big\} \, F^2_{ZZZ} \cr
\noalign{\kern 5pt}
&\quad + 3 \sin^4 \theta_W^0 \big\{ 3 [ (Q_u)^2 T^3_u +  (Q_d)^2 T^3_d  ] + (Q_e)^2 T^3_e  \big\} \, F^3_{ZZZ} \cr
\noalign{\kern 5pt}
&=  0 ~.
\label{anom4}
\end{align}
For $ZWW$,  with Eqs.\  (\ref{ZWWidentity}) and (\ref{Fidentity8}), 
\begin{align}
J_{ZWW}  &=
\big\{ 3 [ T^3_u + T^3_d ] + T^3_\nu +  T^3_e  \big\} \, F^1_{ZWW}  \cr
\noalign{\kern 5pt}
&\quad -  \sin^2 \theta_W^0 \big\{ 3 [ Q_u  +  Q_d] + Q_e\big\} \, F^2_{ZWW} \cr
\noalign{\kern 5pt}
&=  0 ~.
\label{anom5}
\end{align}
Thus all gauge anomalies are canceled in each generation.

So far we have mainly presented the results of gauge anomalies associated with gauge bosons $\gamma, W$ and $Z$.
Generalization to anomalies associated with  KK modes $\gamma^{(n)}, W^{(n)}$ and $Z^{(n)}$ is straightforward.
In the $F$ factors wave functions of gauge bosons are replaced by those of KK gauge bosons.
One comment is in order.  In the GUT-inspired GHU there are dark fermion multiplets listed in Eq.\ (\ref{DFBC1}).
The multiplets $\Psi^{F_q, \alpha}_{({\bf 3},{\bf 4})}$ and $\Psi^{F_\ell , \alpha}_{({\bf 1},{\bf 4})}$  give
contributions to chiral anomalies.  Their contributions are summarized by formulas similar to those presented above.
For instance, the $F^1_{ZZZ}$ part in $J_{ZZZ}$,  contributions from quark-lepton multiplets are 
decomposed as $F^1_{ZZZ} = F^1_{ZZZ} |_{y=0} + F^1_{ZZZ} |_{y=L}$ in Eq.\  (\ref{Fidentity6}).
Contributions from dark fermion multiplets are cast as $F^{1 \, {\rm DF}}_{ZZZ} = - F^1_{ZZZ} |_{y=0} + F^1_{ZZZ} |_{y=L}$.
Thorough discussions of contributions coming from fermion multiplets with various orbifold boundary conditions
in an $SU(2)$ model are given in Ref.\ \cite{AnomalyFlow2}.
Gauge anomalies must be canceled in the dark fermion sector as well.
In the model specified in this paper with boundary conditions (\ref{quarkleptonBC1}) and (\ref{DFBC1}) 
gauge anomalies are canceled both in the quark-lepton and dark fermion sectors.

\section{Summary} 

In the GUT-inspired GHU in the RS space the $W$ and $Z$ couplings of quarks and leptons vary with $\theta_H$.
The magnitude of these couplings depends on fermion species at $\theta_H \not= 0, \pi$.
This fact leads to a question whether or not the cancellation of gauge anomalies is spoiled.
First we noted that all KK modes of quark-lepton multiplets contribute to chiral anomalies.
It has been shown in this paper  that total chiral anomalies including contributions of all KK fermion modes 
are expressed in terms of the values of wave functions of $W$ and $Z$ bosons at the UV and IR branes.
We derived holographic formulas for chiral anomalies.
The total chiral anomalies, which are originally expressed as  sums of products of gauge couplings, 
 are decomposed into products of group factors ($T^3$ and $Q$) and various $F$ factors.
 We showed that $F$ factors are universal and independent of fermion species running  along
 internal triangular loops.
As the $F$ factors are universal, cancellation of anomalies is achieved when 
sums of group factors of fermion multiplets vanish. 
This is indeed what happens in the GUT-inspired GHU model.

In this paper we have focused on gauge anomalies.  The technique developed in deriving holographic formulas
for anomalies is general, and can be applied to global anomalies as well.  It is of great interest to derive
a formula for an anomaly in baryon number current.  In the GUT-inspired GHU in the RS space 
anomaly terms of the form $F_{\mu\nu} \tilde{F}^{\mu\nu}$ contain KK gauge bosons whose anomaly coefficients
are expected to become large for the first KK modes.  It may affect the process of baryon number generation
in the universe.  We hope to come back to this issue in near future.

% \section*{Acknowledgment}
% The author would like to thank % 
 
\vskip 1.cm
 
 \appendix

\section{Basis functions in the RS space} 

Wave functions of gauge and fermion fields in the RS space are expressed in terms of Bessel functions.
For gauge fields we introduce
\begin{align}
 F_{\alpha, \beta}(u, v) &\equiv J_\alpha(u) Y_\beta(v) - Y_\alpha(u) J_\beta(v) ~, \cr
\noalign{\kern 5pt}
 C(z; \lambda) &= \frac{\pi}{2} \lambda z z_L F_{1,0}(\lambda z, \lambda z_L) ~,  \cr
 S(z; \lambda) &= -\frac{\pi}{2} \lambda  z F_{1,1}(\lambda z, \lambda z_L) ~, \cr
 C^\prime (z; \lambda) &= \frac{\pi}{2} \lambda^2 z z_L F_{0,0}(\lambda z, \lambda z_L) ~,  \cr
S^\prime (z; \lambda) &= -\frac{\pi}{2} \lambda^2 z  F_{0,1}(\lambda z, \lambda z_L)~,  \cr
\check S(z; \lambda) &= \frac{C(1; \lambda)}{S(1; \lambda)} \,  S(z; \lambda) ~, 
\label{functionA1}
\end{align}
where $J_\alpha (u)$ and $Y_\alpha (u)$ are Bessel functions of  the first and second kind.
$C(z;\lambda)$ and $S(z; \lambda)$  satisfy
\begin{align}
&- z \frac{d}{dz} \frac{1}{z} \frac{d}{dz} \begin{pmatrix} C \cr S \end{pmatrix} 
= \lambda^{2} \begin{pmatrix} C \cr S \end{pmatrix} ~.
\label{relationA1}
\end{align}
Boundary conditions are given by $C = z_L$, $C'  =S = 0 $, and $S' = \lambda$ at $z=z_L$.
A relation $CS' - S C' = \lambda z$ holds.

For fermion fields with a bulk mass parameter $c$, we define 
\begin{align}
\begin{pmatrix} C_L \cr S_L \end{pmatrix} (z; \lambda,c)
&= \pm \frac{\pi}{2} \lambda \sqrt{z z_L} F_{c+\frac12, c\mp\frac12}(\lambda z, \lambda z_L) ~, \cr
\begin{pmatrix} C_R \cr S_R \end{pmatrix} (z; \lambda,c)
&= \mp \frac{\pi}{2} \lambda \sqrt{z z_L} F_{c- \frac12, c\pm\frac12}(\lambda z, \lambda z_L) ~, \cr
\begin{pmatrix} \check S_L \cr \check C_R \end{pmatrix} (z; \lambda,c)
&=\frac{C_L (1; \lambda, c)}{S_L (1; \lambda, c)} \begin{pmatrix} S_L \cr C_R \end{pmatrix} (z; \lambda,c) ~.
\label{functionA2}
\end{align}
These functions satisfy 
\begin{align}
&D_{+} (c) \begin{pmatrix} C_{L} \cr S_{L} \end{pmatrix} = \lambda  \begin{pmatrix} S_{R} \cr C_{R} \end{pmatrix}, \cr
\noalign{\kern 5pt}
&D_{-} (c) \begin{pmatrix} S_{R} \cr C_{R} \end{pmatrix} = \lambda  \begin{pmatrix} C_{L} \cr S_{L} \end{pmatrix}, \cr
\noalign{\kern 5pt}
&D_{\pm} (c) = \pm \frac{d}{dz} + \frac{c}{z} ~, 
\label{relationA2}
\end{align}
with the boundary conditions $C_{R/L} =1$, $D_-(c) C_R = D_+(c) C_L =S_{R/L} = 0$ at $z=z_{L} $, and 
$C_L C_R - S_L S_R=1$.
For fermion fields with a vector-like mass $m = k \tilde m$,  mode functions are expressed
in terms of
\begin{align}
{\cal C}_{L/R\, 1}(z; \lambda, c, \tilde m) &= C_{L/R} (z; \lambda, c+\tilde{m})+C_{L/R} (z; \lambda, c-\tilde{m}) ~, \cr
{\cal S}_{L/R\, 1}(z; \lambda, c, \tilde m) &= S_{L/R} (z; \lambda, c+\tilde{m})+S_{L/R} (z; \lambda,c-\tilde{m}) ~, \cr
{\cal C}_{L/R\, 2}(z; \lambda, c, \tilde m) &= S_{L/R} (z; \lambda, c+\tilde{m})-S_{L/R} (z; \lambda, c-\tilde{m}) ~, \cr
{\cal S}_{L/R\, 2}(z; \lambda, c, \tilde m) &= C_{L/R} (z; \lambda, c+\tilde{m})-C_{L/R} (z; \lambda, c-\tilde{m}) ~.
\label{MassiveFermion1}
\end{align}

\section{Fermions with $c=0$} 

For fermions with a vanishing bulk mass parameter $c=0$ wave functions are expressed in terms of 
trigonometric functions. Basis functions become
\begin{align}
\begin{pmatrix} C_L \cr S_L \end{pmatrix} (z; \lambda, 0) &= 
\begin{pmatrix} \cos \lambda (z - z_L) \cr  \sin \lambda (z - z_L) \end{pmatrix} , \cr
\noalign{\kern 5pt}
\begin{pmatrix} C_R \cr S_R \end{pmatrix} (z; \lambda, 0) &= 
\begin{pmatrix} \cos \lambda (z - z_L) \cr  - \sin \lambda (z - z_L) \end{pmatrix} .
\label{functionB1}
\end{align}
The spectrum $\{ \lambda_n \}$ determined by Eq.\ (\ref{Up-quark-mass1}) is 
\begin{align}
&\lambda_n = \frac{1}{z_L -1}  \big| n \pi + \onehalf \theta_H \big|  \quad (- \infty < n < \infty) ~.
\label{spectrumB2}
\end{align}
The corresponding wave functions in $1 \le z = e^{ky} \le z_L$ in the original gauge are 
\begin{align}
\begin{pmatrix} f_{L n} (y) \cr g_{L n} (y) \end{pmatrix} 
&= \frac{1}{\sqrt{z_L-1}}  \begin{pmatrix} \cos \alpha_n (z) \cr i \sin \alpha_n (z) \end{pmatrix} , \cr
\noalign{\kern 5pt}
\begin{pmatrix} f_{R n} (y) \cr g_{R n} (y) \end{pmatrix} 
&= \frac{1}{\sqrt{z_L-1}}  \begin{pmatrix} i \sin \alpha_n (z) \cr \cos \alpha_n (z) \end{pmatrix} , \cr
\noalign{\kern 5pt}
\alpha_n (z)  &= \big( n \pi +  \onehalf \theta_H \big) \, \frac{z-z_L}{z_L -1} + \onehalf \theta (z) ~.
\label{waveB3}
\end{align}
Note that $\alpha_n (1) = - n\pi$ and $\alpha_n (z_L) = 0$.
From the boundary conditions
\begin{align}
\begin{pmatrix} f_{L n} (y) \cr g_{L n} (y) \end{pmatrix} 
&= \begin{pmatrix} f_{L n} (- y) \cr - g_{L n} (-y) \end{pmatrix}  
= \begin{pmatrix} f_{L n} (y+ 2L) \cr g_{L n} (y + 2L) \end{pmatrix} , \cr
\noalign{\kern 5pt}
\begin{pmatrix} f_{R n} (y) \cr g_{R n} (y) \end{pmatrix} 
&= \begin{pmatrix} - f_{R n} (-y) \cr g_{R n} (-y) \end{pmatrix} 
= \begin{pmatrix} f_{R n} (y + 2L) \cr g_{R n} (y + 2L) \end{pmatrix} .
\label{boundaryB4}
\end{align}
Making use of Eqs.\ (\ref{waveB3}) and (\ref{boundaryB4}), one finds the relations in Eq.\  (\ref{ABsumrelation1}).

Boundary conditions of dark fermions are given by Eq.\ (\ref{DFBC1}).  For $\Psi^{F_q, \alpha}_{({\bf 3},{\bf 4})} $ and
$\Psi^{F_\ell, \alpha}_{({\bf 1},{\bf 4})}$ with $c_{F_q} = c_{F_\ell} = 0$,   $\{ \lambda_n \}$ and $\alpha_n(z)$ in
Eqs.\ (\ref{spectrumB2}) and (\ref{waveB3}) are replaced by $\{ \lambda_n^{DF} \}$ and $\beta_n(z)$ where
\begin{align}
\lambda_n^{DF} &=  \frac{1}{z_L -1}  \big|  (n + \onehalf ) \pi + \onehalf \theta_H \big|  \quad (- \infty < n < \infty) ~, \cr
\noalign{\kern 5pt}
\beta_n (z) &=  \big( n \pi + \onehalf \pi +  \onehalf \theta_H \big) \, \frac{z-z_L}{z_L -1} + \onehalf \theta (z) ~.
\label{spectrumB5}
\end{align}

\section{Universality of $F$ factors} 

In Section 5 we argued that various $F$ factors appearing in the formulas of chiral anomalies are universal.
For $\gamma \gamma Z^{(\ell)}$ and $\gamma ZZ$ anomalies, for instance, the relations
$F_{Z^{(\ell)}}^{\alpha \, j} = F_{Z^{(\ell)}}^{ j} $ and $F_{\gamma ZZ} ^{\alpha \, k} = F_{\gamma ZZ}^{k}$
$(\alpha = u,d, c, s, t , b , e, \mu, \tau)$  in Eq.\ (\ref{Fidentity1}) are confirmed by showing 
the relations in Eq.\ (\ref{FidentityX1}), 
$F_{Z^{(\ell)}}^{\alpha \, j}  |_{n_\max} \go F_{Z^{(\ell)}}^{ j} $  and 
$F_{\gamma ZZ} ^{\alpha \, k}  |_{n_\max} \go F_{\gamma ZZ}^{k} $ as  $n_\max \go \infty$, 
where $F |_{n_\max}$ is defined by doing the KK sum up to the KK level $n_\max$.
$F |_{n_\max}$ is numerically evaluated. The results for $F_{Z^{(0)}}^{\alpha \, 1}  |_{n_\max}$ 
and $F_{\gamma ZZ}^{\alpha \, 1}  |_{n_\max}$ have been given in Table \ref{Table:ggZ0} and Table \ref{Table:gammaZZ1}
in Section 5, respectively.
In this appendix we present the results for other $F$ factors.

\vskip 10pt
\noindent (i) $F_{Z^{(\ell)}}^{\alpha \, 1} $

The result for $F_{Z^{(0)}}^{\alpha \, 1}  |_{n_\max}$ is given in Table \ref{Table:ggZ0} in Section 5.
The result for $F_{Z^{(1)}}^{\alpha \, 1}  |_{n_\max}$ is given in Table \ref{Table:ggZ1}.  
It converges to $F_{Z^{(1)}}^{1} $ as $n_\max \go \infty$, though the convergence is slower
than in the case of $F_{Z^{(0)}}^{\alpha \, 1}  |_{n_\max}$.

\begin{table}[h]
\renewcommand{\arraystretch}{1.2}
\begin{center}
\caption{$F_{Z^{(1)}}^{\alpha \, 1}  |_{n_\max} $ ($\alpha = u, d, e, t, b, \tau$)  is tabulated
for $\theta_H=0.1$ and $m_\KK = 13\,$TeV.
The bottom row $n_\max = \infty$ represents the value of the formula in Eq.\ (\ref{Fidentity2}).
}
\vskip 10pt
\begin{tabular}{ccccccc}
%\begin{tabular}{|c|c|c|c|c|c|}
\hline 
$n_\max$ & $F_{Z^{(1)}}^{u \, 1}$ & $F_{Z^{(1)}}^{d \, 1}$ &$F_{Z^{(1)}}^{e \, 1}$  
   & $F_{Z^{(1)}}^{t \, 1}$ & $F_{Z^{(1)}}^{b\, 1}$ &$F_{Z^{(1)}}^{\tau \, 1}$  \\
\noalign{\kern 2pt}
\hline
0  &$5.56610$ &$5.56610$ &$5.74876$  &$4.40691$ &$4.41348$ &$5.29107$ \\
10 &$3.84398$ &$3.84398$ &$3.89273$ &$3.63898$ &$3.63901$ &$3.78169$ \\
20 &$3.70283$ &$3.70283$ &$3.73028$ &$3.59060$ &$3.59062$ &$3.66819$\\
30 &$3.65039$ &$3.65039$ &$3.66949$ &$3.57312$ &$3.57315$ &$3.62640$\ \\
40 &$3.62303$  &$3.62303$ &$3.63766$ &$3.56411$ &$ 3.56413$ &$3.60468$\\
\noalign{\kern 5pt}
\hline
$\infty$ &  & &$3.53591$ & && \\
\hline 
\end{tabular}
\label{Table:ggZ1}
\end{center}
\end{table}

\vskip 10pt
\noindent (ii) $F_{\gamma ZZ}^{\alpha \, 2}$ and $F_{\gamma WW}^{\alpha}$

The result for $F_{\gamma ZZ}^{\alpha \, 1}  |_{n_\max}$ ($\alpha = u, d, e, t, b, \tau$) is given in Table \ref{Table:gammaZZ1}
in Section 5.  
The result for $F_{\gamma ZZ}^{\alpha \, 2}  |_{n_\max}$ is given in Table \ref{Table:gammaZZ2}.
The result for $F_{\gamma WW}^{\alpha}  |_{n_\max} $ ($\alpha = ud, \nu_e e, tb, \nu_\tau \tau$)  is
tabulated in Table \ref{Table:gammaWW}.

\begin{table}[h]
\renewcommand{\arraystretch}{1.2}
\begin{center}
\caption{$F_{\gamma ZZ}^{\alpha \, 2}  |_{n_\max} $ ($\alpha = u, d, e, t, b, \tau$)  is tabulated
for $\theta_H=0.1$ and $m_\KK = 13\,$TeV.
The bottom row $n_\max = \infty$ represents the value of the formula in Eq.\ (\ref{Fidentity3}).
}
\vskip 10pt
\begin{tabular}{ccccccc}
%\begin{tabular}{|c|c|c|c|c|c|}
\hline 
$n_\max$ & $F_{\gamma ZZ}^{u \, 2}$ & $F_{\gamma ZZ}^{d \, 2}$ &$F_{\gamma ZZ}^{e \, 2}$  
   & $F_{\gamma ZZ}^{t \, 2}$ & $F_{\gamma ZZ}^{b\, 2}$ &$F_{\gamma ZZ}^{\tau \, 2}$  \\
\noalign{\kern 2pt}
\hline
0  &$0.997874$ &$0.997874$ &$0.997879$  &$0.997841$ &$0.997841$ &$0.997866$ \\
6 &$0.997817$ &$0.997817$ &$0.997820$ &$0.997804$ &$0.997804$ &$0.997813$ \\
12&$0.997808$ &$0.997808$ &$0.997810$ &$0.997800$ &$0.997800$ &$0.997805$\\
18 &$0.997805$ &$0.997805,$ &$0.997806$ &$0.997799$ &$0.997799$ &$0.997802$\ \\
24 &$0.997803$  &$0.997803$ &$0.997804$ &$0.997798$ &$0.997798$ &$0.997800$\\
\noalign{\kern 5pt}
\hline
$\infty$ &  & &$0.997796$ & && \\
\hline 
\end{tabular}
\label{Table:gammaZZ2}
\end{center}
\end{table}

\begin{table}[h]
\renewcommand{\arraystretch}{1.2}
\begin{center}
\caption{$F_{\gamma WW}^{\alpha}  |_{n_\max} $ ($\alpha = ud, \nu_e e, tb, \nu_\tau \tau$)  is tabulated
for $\theta_H=0.1$ and $m_\KK = 13\,$TeV.
The bottom row $n_\max = \infty$ represents the value of the formula in Eq.\ (\ref{Fidentity5}).
}
\vskip 10pt
\begin{tabular}{ccccc}
%\begin{tabular}{|c|c|c|c|c|c|}
\hline 
$n_\max$ & $F_{\gamma WW}^{ud}$ & $F_{\gamma WW}^{\nu_e e}$ &$F_{\gamma WW}^{tb}$  
   & $F_{\gamma WW}^{ \nu_\tau \tau}$   \\
\noalign{\kern 2pt}
\hline
0  &$0.995296$ &$0.995300$ &$0.995942$  &$0.995290$  \\
6 &$0.997407$ &$0.997360$ &$0.997616$ &$0.997468$  \\
12&$ $0.997558&$0.997531$ &$0.997669$ &$0.997592$ \\
18 &$0.997612$ &$0.997593$ &$0.997688$ &$0.997636$  \\
24 &$0.997640$  &$0.997626$ &$0.997697$ &$0.997658$ \\
\noalign{\kern 5pt}
\hline
$\infty$ &   &$0.997728$ & & \\
\hline 
\end{tabular}
\label{Table:gammaWW}
\end{center}
\end{table}

\vskip 10pt
\noindent (iii) $F_{ZZZ}^{\alpha \, 1}$, $F_{ZZZ}^{\alpha \, 2}$ and $F_{ZZZ}^{\alpha \, 4}$

The result for $F_{ZZZ}^{\alpha \, 1}  |_{n_\max} $ ($\alpha = u, d, \nu_e, e, t, b, \nu_\tau, \tau$) is given in 
Table \ref{Table:ZZZ1}.   $F_{ZZZ}^{u\, 1}  |_{n_\max} \sim F_{ZZZ}^{d\, 1}  |_{n_\max} $ and 
$F_{ZZZ}^{\nu_e \, 1}  |_{n_\max} \sim F_{ZZZ}^{e \, 1}  |_{n_\max} $ in six-digit accuracy.
The result for $F_{ZZZ}^{\alpha \, 2}  |_{n_\max} $ ($\alpha = u, d,  e, t, b, \tau$)  
is tabulated in Table \ref{Table:ZZZ2}.
$F_{ZZZ}^{u\, 2}  |_{n_\max} \sim F_{ZZZ}^{d\, 2}  |_{n_\max} $  in six-digit accuracy.

The result for $F_{ZZZ}^{\alpha \, 4}  |_{n_\max} $ ($\alpha = u, d, e, t, b,  \tau$)  is given in Table  \ref{Table:ZZZ4}.
As remarked in the text, 
in evaluating $F_{ZZZ}^{\alpha \, 4}  |_{n_\max} $ ($\alpha = d,  s, b$)  
contributions coming from $D_d, D_s, D_b$ towers are very important.
For instance, for $n_\max = 24$,  $F_{ZZZ}^{b \, 4} |_{b \, {\rm tower} \,  {\rm loop}}= - 0.000188$ whereas 
$F_{ZZZ}^{b \, 4} |_{D_b \, {\rm tower} \,  {\rm loop}}= 0.000179$.

\begin{table}[h]
\renewcommand{\arraystretch}{1.2}
\begin{center}
\caption{$F_{ZZZ}^{\alpha \, 1}  |_{n_\max} $ ($\alpha = u, d, \nu_e, e, t, b, \nu_\tau, \tau$)  is tabulated
for $\theta_H=0.1$ and $m_\KK = 13\,$TeV.
The bottom row $n_\max = \infty$ represents the value of the formula in Eq.\ (\ref{Fidentity6}).
}
\vskip 10pt
\begin{tabular}{ccccccc}
%\begin{tabular}{|c|c|c|c|c|c|}
\hline 
$n_\max$ & $F_{ZZZ}^{u \, 1} ,F_{ZZZ}^{d \, 1}$ & $F_{ZZZ}^{\nu_e \, 1}  , F_{ZZZ}^{e \, 1}$
 &$F_{ZZZ}^{t\, 1}$     & $F_{ZZZ}^{b\, 1}$ & $F_{ZZZ}^{\nu_\tau\, 1}$ &$F_{ZZZ}^{\tau \, 1}$  \\
\noalign{\kern 2pt}
\hline
0  &$0.993080$ &$  0.993088$ &$ 0.995041$ &$ 0.993030$ &$ 0.993069$ &$0.993069$  \\
6 &$0.996227$ &$ 0.996158$ &$0.996547$ &$0.996527$ &$ 0.996318$ &$0.996317$  \\
12&$0.996450$ &$0.996410$ &$0.996621$ &$0.996610$ &$0.996502$ &$0.996500$ \\
18 &$0.996531$ &$0.996503$ &$0.996648$ &$0.996638$ &$0.996567$ &$ 0.996565$  \\
24 &$0.996573$ &$0.996551$ &$ 0.996661$ &$0.996652$ &$0.996600$ &$0.996598$  \\
\noalign{\kern 5pt}
\hline
$\infty$ &  & &$0.996703$ & && \\
\hline 
\end{tabular}
\label{Table:ZZZ1}
\end{center}
\end{table}

\begin{table}[h]
\renewcommand{\arraystretch}{1.2}
\begin{center}
\caption{$F_{ZZZ}^{\alpha \, 2}  |_{n_\max} $ ($\alpha = u, d, , e, t, b,  \tau$)  is tabulated
for $\theta_H=0.1$ and $m_\KK = 13\,$TeV.
The bottom row $n_\max = \infty$ represents the value of the formula in Eq.\ (\ref{Fidentity7}).
}
\vskip 10pt
\begin{tabular}{cccccc}
%\begin{tabular}{|c|c|c|c|c|c|}
\hline 
$n_\max$ & $F_{ZZZ}^{u \, 2} ,F_{ZZZ}^{d \, 2}$ & $F_{ZZZ}^{e \, 2}$
 &$F_{ZZZ}^{t\, 2}$   & $F_{ZZZ}^{b\, 2}$ & $F_{ZZZ}^{\tau\, 2}$  \\
\noalign{\kern 2pt}
\hline
0  &$0.995567$ &$ 0.995574$ &$0.996860$ &$0.995517$ &$0.995556$  \\
6 &$0.997637$ &$0.997592$ &$ 0.997843$ &$0.997830$ &$0.997695$  \\
12&$0.997781$ &$ 0.997755$ &$0.997891$ &$0.997883$ &$0.997813$  \\
18 &$0.997833$ &$0.997815 $&$0.997908$ &$0.997901$ &$0.997855$  \\
24 &$0.997860$ &$0.997846$ &$0.997916$ &$0.997910$ &$0.997876$  \\
\noalign{\kern 5pt}
\hline
$\infty$ &  & &$0.997943$ & & \\
\hline 
\end{tabular}
\label{Table:ZZZ2}
\end{center}
\end{table}

\begin{table}[h]
\renewcommand{\arraystretch}{1.2}
\begin{center}
\caption{$F_{ZZZ}^{\alpha \, 4}  |_{n_\max} $ ($\alpha = u, d, e, t, b,  \tau$)  is tabulated
for $\theta_H=0.1$ and $m_\KK = 13\,$TeV.
The bottom row $n_\max = \infty$ represents the value of the formula in Eq.\ (\ref{Fidentity7}).
}
\vskip 10pt
\begin{tabular}{ccccccc}
%\begin{tabular}{|c|c|c|c|c|c|}
\hline 
$n_\max$ & $F_{ZZZ}^{u \, 4} $ & $F_{ZZZ}^{d\, 4} $
 &$F_{ZZZ}^{e\, 4}$     & $F_{ZZZ}^{t\, 4}$ & $F_{ZZZ}^{b \, 4}$ &$F_{ZZZ}^{\tau \, 4}$  \\
\noalign{\kern 2pt}
\hline
0  &$0.000299$ &$ 0.000209$ &$0.000306$ &$0.000127$ &$0.000054$ &$0.000287$  \\
6 &$0.000066$ &$0.000034$ &$0.000075$ &$0.000023$ &$ -0.000026$ &$0.000054$  \\
12&$0.000037$ &$0.000019$ &$0.000043$ &$0.000012$ &$-0.000016$ &$ 0.000030$ \\
18 &$0.000026$ &$0.000014$ &$0.000030$ &$0.000009$ &$-0.000011$ &$0.000021$ \\
24 &$0.000020$ &$0.000011$ &$0.000023$ &$0.000006$ &$-0.000009$ &$ 0.000016$ \\
\noalign{\kern 5pt}
\hline
$\infty$ &  & &$0$ & && \\
\hline 
\end{tabular}
\label{Table:ZZZ4}
\end{center}
\end{table}

\vskip 10pt
\noindent (iv) $F_{ZWW}^{\alpha \, 1}$ and  $F_{ZWW}^{\alpha \, 2}$ 

The result for $F_{ZWW}^{\alpha \, 1}  |_{n_\max} $ ($\alpha = u, d, \nu_e, e, t, b, \nu_\tau, \tau$)  is given in 
Table \ref{Table:ZWW1}.    $F_{ZWW}^{u\, 1}  |_{n_\max} \sim F_{ZWW}^{d\, 1}  |_{n_\max} $ and 
$F_{ZWW}^{\nu_e \, 1}  |_{n_\max} \sim F_{ZWW}^{e \, 1}  |_{n_\max} $ in six-digit accuracy.
The result for $F_{ZWW}^{\alpha \, 2}  |_{n_\max} $ ($\alpha = u, d, , e, t, b,  \tau$)   is given in 
Table \ref{Table:ZWW2}.   $F_{ZWW}^{u\, 2}  |_{n_\max} \sim F_{ZWW}^{d\, 2}  |_{n_\max} $ and
$F_{ZWW}^{t\, 2}  |_{n_\max} \sim F_{ZWW}^{b\, 2}  |_{n_\max} $ in six-digit accuracy.

\begin{table}[h]
\renewcommand{\arraystretch}{1.2}
\begin{center}
\caption{$F_{ZWW}^{\alpha \, 1}  |_{n_\max} $ ($\alpha = u, d, \nu_e, e, t, b, \nu_\tau, \tau$)  is tabulated
for $\theta_H=0.1$ and $m_\KK = 13\,$TeV.
The bottom row $n_\max = \infty$ represents the value of the formula in Eq.\ (\ref{Fidentity8}).
}
\vskip 10pt
\begin{tabular}{ccccccc}
%\begin{tabular}{|c|c|c|c|c|c|}
\hline 
$n_\max$ & $F_{ZWW}^{u \, 1} ,F_{ZWW}^{d \, 1}$ & $F_{ZWW}^{\nu_e \, 1}  , F_{ZWW}^{e \, 1}$
 &$F_{ZWW}^{t\, 1}$     & $F_{ZWW}^{b\, 1}$ & $F_{ZWW}^{\nu_\tau\, 1}$ &$F_{ZWW}^{\tau \, 2}$  \\
\noalign{\kern 2pt}
\hline
0  &$0.992995$ &$ 0.993001$ &$0.994292$ &$0.993622$ &$0.992985$ &$0.992985$   \\
6 &$0.996155$ &$0.996085$ &$0.996471$ &$0.996464$ &$0.996247$ &$0.996246$   \\
12&$0.996380$ &$0.996340$ &$0.996549$ &$0.996545$ &$0.996431$ &$0.996431$  \\
18 &$0.996462$ &$0.996433$ &$0.996576$ &$0.996573$ &$0.996497$ &$0.996496$   \\
24 &$0.996504$ &$0.996482$ &$0.996590$ &$0.996587$ &$ 0.996530$ &$0.996530$  \\
\noalign{\kern 5pt}
\hline
$\infty$ &  & &$0.996635$ & && \\
\hline 
\end{tabular}
\label{Table:ZWW1}
\end{center}
\end{table}

\begin{table}[h]
\renewcommand{\arraystretch}{1.2}
\begin{center}
\caption{$F_{ZWW}^{\alpha \, 2}  |_{n_\max} $ ($\alpha = u, d, , e, t, b,  \tau$)  is tabulated
for $\theta_H=0.1$ and $m_\KK = 13\,$TeV.
The bottom row $n_\max = \infty$ represents the value of the formula in Eq.\ (\ref{Fidentity8}).
}
\vskip 10pt
\begin{tabular}{ccccc}
%\begin{tabular}{|c|c|c|c|c|c|}
\hline 
$n_\max$ & $F_{ZWW}^{u \, 2} ,F_{ZWW}^{d \, 2}$ & $F_{ZWW}^{e \, 2}$
 &$F_{ZWW}^{t\, 2}, F_{ZWW}^{b\, 2}$    & $F_{ZWW}^{\tau\, 2}$  \\
\noalign{\kern 2pt}
\hline
0  &$0.995482$ &$ 0.995488$ &$ 0.996111$ &$ 0.995472$  \\
6 &$0.997564$ &$ 0.997519$ &$ 0.997767$ &$ 0.997624$ \\
12&$0.997711$ &$0.997684$ &$0.997819$ &$0.997744$  \\
18 &$0.997763$ &$ 0.997745$ &$0.997836$ &$0.997786$ \\
24 &$0.997791$ &$0.997777$ &$ 0.997845$ &$0.997808$  \\
\noalign{\kern 5pt}
\hline
$\infty$ &  & $0.997876$ &&  \\
\hline 
\end{tabular}
\label{Table:ZWW2}
\end{center}
\end{table}

\vskip 1.cm

%\newpage
 
 %%%%%%%. references  %%%%%%%%%%%

% A useful Journal macro
%\def\jnl#1#2#3#4{{#1}{\bf #2} (#4) #3}
\def\jnl#1#2#3#4{{#1}{\bf #2},  #3 (#4)}

\def\Zphys{{\em Z.\ Phys.} }
\def\jssc{{\em J.\ Solid State Chem.\ }}
\def\jpsJ{{\em J.\ Phys.\ Soc.\ Japan }}
\def\ptps{{\em Prog.\ Theoret.\ Phys.\ Suppl.\ }}
\def\PTP{{\em Prog.\ Theoret.\ Phys.\  }}
\def\PTEP{{\em Prog.\ Theoret.\ Exp.\  Phys.\  }}
\def\JMP{{\em J. Math.\ Phys.} }
\def\NPB{{\em Nucl.\ Phys.} B}
\def\NP{{\em Nucl.\ Phys.} }
\def\PLB{{\it Phys.\ Lett.} B}
\def\PL{{\em Phys.\ Lett.} }
\def\PRL{\em Phys.\ Rev.\ Lett. }
\def\PRB{{\em Phys.\ Rev.} B}
\def\PRD{{\em Phys.\ Rev.} D}
\def\PRe{{\em Phys.\ Rep.} }
\def\AP{{\em Ann.\ Phys.\ (N.Y.)} }
\def\RMP{{\em Rev.\ Mod.\ Phys.} }
\def\ZPC{{\em Z.\ Phys.} C}
\def\SCI{{\em Science} }
\def\CMP{\em Comm.\ Math.\ Phys. }
\def\MPLA{{\em Mod.\ Phys.\ Lett.} A}
\def\IJMPA{{\em Int.\ J.\ Mod.\ Phys.} A}
\def\IJMPB{{\em Int.\ J.\ Mod.\ Phys.} B}
\def\EPJC{{\em Eur.\ Phys.\ J.} C}
\def\EPJP{{\em Eur.\ Phys.\ J.} Plus}
\def\PR{{\em Phys.\ Rev.} }
\def\JHEP{{\em JHEP} }
\def\JCAP{{\em JCAP} }
\def\cmp{{\em Com.\ Math.\ Phys.}}
\def\JPA{{\em J.\  Phys.} A}
\def\JPG{{\em J.\  Phys.} G}
\def\NJP{{\em New.\ J.\  Phys.} }
\def\CQG{\em Class.\ Quant.\ Grav. }
\def\ATMP{{\em Adv.\ Theoret.\ Math.\ Phys.} }
\def\ibid{{\em ibid.} }
\def\ChP{{\em Chin.Phys.}C}
\def\NCA{{\it Nuovo Cim.} A}

\renewenvironment{thebibliography}[1]
         {\begin{list}{[$\,$\arabic{enumi}$\,$]}  % {\arabic{enumi}.}
         {\usecounter{enumi}\setlength{\parsep}{0pt}
          \setlength{\itemsep}{0pt}  \renewcommand{\baselinestretch}{1.2}
          \settowidth
         {\labelwidth}{#1 ~ ~}\sloppy}}{\end{list}}

\newpage
 \vskip 1.cm

\leftline{\Large \bf References}

%%%%%%%%%%%%% BIBLIOGRAPHY  %%%%%%%%%%%%%%%%%%%%

\end{document}